%% file: main.tex
\documentclass[journal]{IEEEtran}
\usepackage{amsmath,amsfonts}
\usepackage{array}
\usepackage[caption=false,font=normalsize,labelfont=sf,textfont=sf]{subfig}
\usepackage{textcomp}
\usepackage{stfloats}
\usepackage{url}
\usepackage{verbatim}
\usepackage{graphicx}
\input{misc/customize}

\usepackage{algorithm}
\usepackage{algpseudocode} 
\usepackage{enumitem}
\usepackage{xcolor}
\usepackage[
    colorlinks=true,
    linkcolor=blue,   % internal links: \ref, \eqref, etc.
    citecolor=blue,   % citation links
    urlcolor=blue     % URLs
]{hyperref}

\def\BibTeX{{\rm B\kern-.05em{\sc i\kern-.025em b}\kern-.08em
    T\kern-.1667em\lower.7ex\hbox{E}\kern-.125emX}}
\usepackage{balance}
\begin{document}
\bstctlcite{IEEEexample:BSTcontrol}
\title{Row-Based Layout Synthesis for Analog Circuits\\Using Height-Quantized Primitives}
\author{Endalk Y. Gebru, Ramprasath S., Ramesh Harjani, and Sachin S. Sapatnekar 

\thanks{E. Y. Gebru, R. Harjani, and S. S. Sapatnekar are with the Department of Electrical and Computer Engineering at the University of Minnesota, Minneapolis, MN, USA. Ramprasath S. is with the Department of Electrical Engineering at the Indian Institute of Technology, Madras, Chennai, India.}}

\markboth{IEEE Transactions on Computer-Aided Design of Integrated Circuits and Systems,~Vol.~X, No.~Y, Month~Year}%
{Row-Based Layout Synthesis for Analog Circuits using Height-Quantized Primitives}

\maketitle

\input{contents/1_Abstract}

\input{contents/2_Introduction}
\input{contents/3_Quantized_height_layout_synthesis}
\input{contents/4_Row_based_layout_synthesis}
\input{contents/5_Results}

\input{contents/6_Conclusion}

\section*{Acknowledgments}
\noindent
This work is supported in part by the National Science Foundation, award 2212345. ChatGPT was used for language polishing and presentation clarity. All technical content, results, and conclusions were reviewed and verified by the authors.

\bibliographystyle{IEEEtran}
\bibliography{bib/venue_abbrev, bib/main}
\end{document}

%% file: misc/customize.tex
\usepackage{float}
\usepackage{tabularx}
\usepackage{mathtools}
\usepackage{amsmath}
\usepackage{hhline}
\usepackage{silence}
\usepackage{graphicx}
\usepackage{soul}
\usepackage{multirow}
\usepackage{graphicx}
\usepackage{multirow}
\usepackage{amsmath}
\usepackage[english]{babel}
\usepackage{booktabs}
\usepackage{array}
\usepackage{paralist}
\usepackage{threeparttable}
\usepackage{lipsum}
\usepackage{flushend}
\usepackage{cuted}
\usepackage{algpseudocode}
\usepackage{algorithm}
\usepackage{amssymb}
\usepackage{multicol}
\usepackage{subfig}
\usepackage{cite}
\usepackage{makecell}
\usepackage{url}
\usepackage[table]{xcolor}
\usepackage[normalem]{ulem}

\newcommand{\ignore}[1]{}

\definecolor{gray1}{gray}{0.7}

\algnewcommand{\LeftComment}[1]{\(\triangleright\) #1}

\usepackage{listings}

%% file: contents/1_Abstract.tex
\begin{abstract}
Restrictive design rules and strong layout-dependent effects have tightened the coupling between physical layout decisions and electrical performance in advanced process nodes, such as FinFET, making analog and mixed-signal (AMS) layout automation increasingly difficult. This paper presents a quantized row-height layout synthesis methodology for AMS circuits, a methodology that has previously been shown to reduce the simulation-to-silicon gap. The proposed flow optimizes a row-height fabric from circuit requirements and layout constraints while mapping analog building blocks into quantized-height rows. Results on multiple testcases demonstrate that the proposed flow synthesizes layouts with similar postlayout performance relative to less-constrained custom baseline designs, with comparable performance metrics. Our quantized-height designs are shown to reduce the schematic-to-postlayout performance gap by up to $\mathbf{68.5\%}$ and result in lower area for most of our testcases, with a maximum area reduction of $\mathbf{24.1\%}$.
\end{abstract}

%% file: contents/2_Introduction.tex
\section{Introduction}
\noindent
The progression to advanced FinFET and nanosheet nodes has reshaped the physical design fabric. As compared to planar technologies, the design space for analog and mixed-signal (AMS) layout is far more constrained and quantized.  Fixed-pitch device geometries, unidirectional routing metals, and restrictive patterning constraints amplify layout-dependent and local-layout-dependent effects (LDE/LLEs) and increase sensitivity to discrete implementation choices~\cite{Wang22, Rossoni25, Dhar21}. Circuit performance depends not only on schematic parameters but also on layout choices such as device folding, fin allocation, and routing~\cite{Lewyn09}. These effects must be considered during physical design as optimized schematics based purely on prelayout estimates of parasitics exhibit large postlayout performance deviation due to parasitic- and LDE-induced shifts~\cite{Ou15}, widening the simulation-to-silicon (S2S) gap~\cite{Chou25, Lewyn09}.

Existing automated AMS layout frameworks~\cite{Kunal19, Dhar21, Chen21, Xu19} have shown that constraint-driven placement and routing can operate within these restrictive fabrics and enforce key analog layout requirements such as symmetry and matching. However, these approaches remain primarily focused on layout feasibility and constraint satisfaction, often treating parasitic- and LDE-induced degradation as a postlayout problem rather than an integral part of performance optimization. 

In practice, existing AMS physical-design methodologies can be broadly viewed through three dominant paradigms: (i)~full-custom manual layout remains the strongest baseline for the best quality; however, its heavy dependency on designer effort limits scalability, productivity, and repeatability across circuits and technology nodes \cite{Scheible22}; (ii)~analog-targeted automation frameworks such as ALIGN~\cite{Kunal19, Dhar21} and MAGICAL~\cite{Chen21, Xu19} represent a scalable AMS physical-design direction by automating constraint-driven placement and routing for a flexible design flow; (iii)~more recent structured library- and template-based methodologies improve regularity and productivity by constraining implementation to predefined, digital-standard-cell-like layout abstractions with well-defined physical interfaces and fixed row heights: these methods are increasingly seen as a promising path for scalable AMS design. For example, an approach for building analog standard cells for field-programmable analog arrays using floating-gate technology is presented in~\cite{Hasler20, Hasler24}.  For more mainstream analog design, ALOE/Open-ALOE~\cite{Wei21,Li25} leverage structured analog building blocks called stem cells with commercial place-and-route (PnR) engines for digital layout, and abuttable analog cell (Acell) libraries are proposed in~\cite{Zhou25}. The use of fixed-height row-based layouts is even more important in advanced technology nodes with restrictive design rules: a flow based on abuttable, fixed-height cells in~\cite{Chou25} that shows improved layout uniformity makes a convincing case that such a methodology is imperative for reducing the S2S gap, and is also friendlier for layout migration across technology nodes.

\begin{figure}[t]
\centering
\includegraphics[width=1\linewidth]{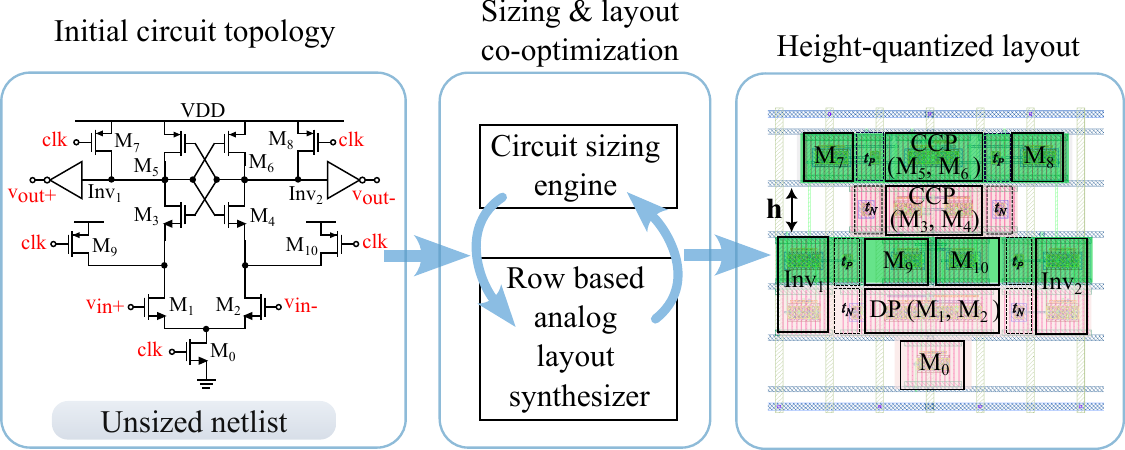}\caption{Proposed height-quantized AMS flow from an unsized netlist to a final row-based layout through circuit sizing and layout synthesis, illustrated using a StrongARM comparator example.}
\label{fig:fixed-height-arch}
\end{figure}

This work proposes a structured fixed-height row-based analog layout synthesis approach for FinFET technologies. Our approach is illustrated in Figure~\ref{fig:fixed-height-arch}: the figure shows the schematic of a StrongARM comparator and its corresponding row-based layout produced by our approach, where each cell in the layout corresponds to a \textit{primitive}, i.e., a functional unit consisting of a small set of transistors~\cite{Kunal19}. This fixed-height organization improves layout regularity, pin accessibility, and placement scalability by mapping analog building blocks into a standard-cell-like row structure. At the same time, analog layout cannot be reduced to purely geometric packing: symmetry, proximity, well organization, and substrate-tie access remain important physical constraints that must be considered during placement. Among these constraints, well and tap planning provide a representative example of the coupling between circuit layout regularity and physical implementation requirements. Insufficient tap access can degrade substrate-bias robustness, whereas excessive tap insertion increases area and may reduce placement compactness. Prior works have addressed related aspects of well generation and tap insertion, including designer-like well-shape generation after placement~\cite{Xu19_2} and graph-based tap insertion with coverage and symmetry constraints~\cite{Ramprasath23}. This work treats well and tap planning as part of the row-based placement and refinement problem rather than as a detached layout-cleanup step. Our flow combines fixed-height primitive realization, well-aware placement, tap-aware refinement, and compact row-based assembly in a unified analog-aware synthesis framework.

The main contributions of this work are summarized below, and a comparison with prior methods is shown in Table~\ref{tab:related_work}.

\noindent
\textbf{Primitive-based synthesis.} Our approach is based on hierarchical layout synthesis where the lowest level of hierarchy is a primitive. In contrast with prior approaches~\cite{Hasler20, Hasler24,Li25,Wei21,Chou25,Zhou25} where the unit or stem cell is a transistor, our method naturally addresses primitive-level matching and symmetry, e.g., within a differential pair (DP) or a cross-coupled pair (CCP), as shown in Figure~\ref{fig:fixed-height-arch}. This primitive abstraction also reduces the number of analog devices exposed to the top-level place-and-route flows by encapsulating deterministic device arrangements into reusable primitives, thereby easing the burden on the main placement and routing stages.

\noindent
\textbf{Performance-driven primitive derivation with multilayer-perceptron (MLP)-assisted height optimization.} Our approach dynamically determines the geometry of primitive cells and optimizes the common cell height using MLP-based surrogate models that map electrical performance metrics into layout-level height-selection decisions while accounting for FinFET-specific layout constraints. Moreover, prior methods \cite{Chou25, Hasler20, Hasler24} constrain device realization to predefined geometries and a very limited and fixed set of folding options from predefined libraries and restrict circuit-specific co-optimization of fin allocation, multifingering, and stacking, all of which are incorporated in our approach.
For example, our technique achieves a $\sim10\%$ improvement in figure of merit for a five-transistor  operational transconductance amplifier (OTA), compared with layouts using arbitrary, unoptimized fixed cell heights as in~\cite{Hasler20,Hasler24,Wei21,Li25,Chou25,Zhou25}.

\noindent
\textbf{Analog-aware fixed-height PnR framework.} Prior methods~\cite{Hasler20,Hasler24,Wei21,Li25} use digital PnR engines to improve layout scalability, but analog-specific constraints such as symmetry, matching, well continuity, and tap distributions are not comprehensively addressed. Existing well- and tap-aware methods~\cite{Xu19_2, Ramprasath23} either generate well shapes as a postplacement step without jointly optimizing tap selection, or formulate tap insertion around explicitly specified local tap relations rather than deriving tap requirements from the placement-induced
well-island structure and a physically motivated conductance-based model of substrate/well coverage.
To address these limitations, we present a customized automated PnR engine tailored for analog layouts with fixed cell heights. The placement problem is formulated as a mixed-integer linear program (MILP) that incorporates considerations related to matching/symmetry and well definition -- specifically, maximizing well-continuity and optimizing tap insertion. Figure~\ref{fig:fixed-height-arch} illustrates the well clustering result, where the green regions denote the N-well and the red regions denote the P-well region. The corresponding well taps, P- and N-taps denoted by $t_P$ and $t_N$, respectively, are optimized within each clustered well using a cluster-restricted tap-coverage model based on an inverse-distance conductance proxy, as described in detail in Section~\ref{sec:layout_synthesis}. This formulation selects a compact set of taps that satisfies local substrate/well coverage requirements, avoiding unnecessary tap insertion.

\input{tables/table_1_prior_work_comparison}

The paper is organized as follows. Section~\ref{sec:framework} overviews the optimization framework and the sizing methodology that drives the schematic-layout iteration loop. Next, the complete quantized-height layout synthesis methodology, which forms the core of the paper, is detailed in Section~\ref{sec:layout_synthesis}.  Experimental results on various circuit topologies are then described in Section~\ref{sec:results}, followed by concluding remarks in Section~\ref{sec:conclusion}.

%% file: tables/table_1_prior_work_comparison.tex
\newcommand{\cYes}{\textbf{\checkmark}}
\newcommand{\cNo}{$\textsf{x}$} 
\newcommand{\cNS}{N/A}
\begin{table}[t]
\centering
\caption{Comparison of our method against related works.}
\label{tab:related_work}
\renewcommand{\arraystretch}{1.0}
\footnotesize 
\begin{tabular}{|p{0.36\columnwidth}|c | c | c | c|}
\hline
\textbf{Features} & \textbf{\cite{Chou25}} &\textbf{\cite{Wei21,Li25}} & \textbf{\cite{Zhou25}} & \textbf{This work} \\
\hline
\hline
Primitive-based synthesis   & \cNo      & \cNo      & \cYes     & \cYes  \\ \hline
Sizing-layout integration   & \cNo      & \cNo      & \cNo      & \cYes  \\ \hline
% Automated height optimization         & \cNo      & \cNo      & \cNo      & \cYes  \\ \hline
Placement/routing           & --        & Digital   & Digital   & Custom \\ \hline
Device symmetry/matching    & \cNo      & \cNo      & \cYes     & \cYes  \\ \hline
Tap Optimization            & \cNo      & \cNo      & \cNo      & \cYes  \\ \hline
Well-island clustering      & \cNo      & \cNo      & \cNo      & \cYes  \\ \hline
\end{tabular}
\end{table}

%% file: contents/3_Quantized_height_layout_synthesis.tex
\section{Framework Overview and Sizing Front End}
\label{sec:framework}

\begin{figure}[t]
    \centering
\includegraphics[width=1\linewidth]{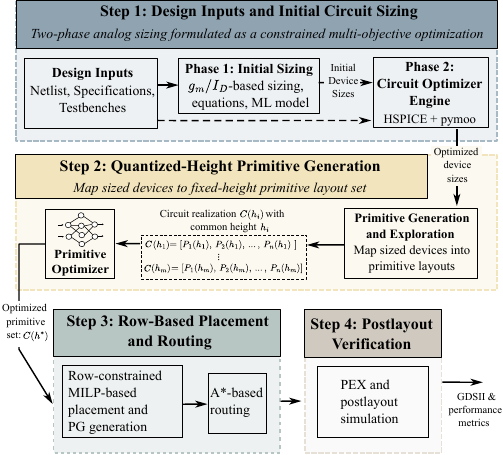}
    \caption{Proposed design flow. In Step 2, $C(h_i)$ denotes the circuit realization at common height $h_i$, where $P_j(h_i)$ is primitive $j$ at height $h_i$.}
    \label{fig:flowchart}
\end{figure}

\noindent
The proposed methodology, summarized in Figure~\ref{fig:flowchart}, goes from circuit-level specifications to quantized row-based layout generation in four steps: (1)~Given the circuit netlist, target electrical specifications, and initial sizing scripts, the framework performs a front-end circuit sizing procedure based on~\cite{Gebru26}. Next, it proceeds to (2)~fixed-height primitive generation and selection, followed by (3)~row-based placement and routing, and finally (4)~post-layout extraction and simulation.

\noindent
\textbf{Step 1: Design Inputs and Front-end Circuit Sizing.}
The flow starts from the circuit netlist, target specifications, and technology constraints, and first solves an initial sizing problem to obtain a feasible circuit implementation. In general, this stage can be viewed as a simulation-driven constrained optimization problem over the design variables $\mathbf{x} \in \mathcal{X}$, with circuit performance being evaluated with SPICE simulations and compared against target specifications. Let $\mathbf{g}(\mathbf{x})=[g_1(\mathbf{x}),g_2(\mathbf{x}),\ldots,g_m(\mathbf{x})]^T$ denote the vector of schematic-level simulated performance metrics, where $g_i(\mathbf{x})$ is the simulated value of the $i^{\rm th}$ metric, and let $\mathbf{S}=[S_1,S_2,\ldots,S_m]^T$ denote the corresponding target specification.

The front-end sizing step, based on~\cite{Gebru26}, is outlined here for completeness.  It adopts a two-phase sizing procedure.
In the first phase, topology-dependent analytical scripts together with the $g_m/I_D$ method \cite{Silveira96, Jespers17, Gebru26} are used to construct a circuit-aware initial solution $\mathbf{x}_0$, thereby restricting the starting point to an admissible operating region. For circuits without an initialization script, the flow also supports direct optimization by passing the design inputs directly to the Phase~2 optimization engine, as shown by the dotted arrow in Figure~\ref{fig:flowchart}. In the second phase, this initial solution is refined through bounded simulation-in-the-loop optimization using derivative-free optimizers such as Differential Evolution (DE)~\cite{storn_97} with HSPICE in the loop. For scalarized refinement using DE, multiple design specifications are mapped into a unified weighted cost function to be minimized:
\begin{align}
    C_{DE}(\mathbf{x})& =\sum_{i=1}^{m}w_i\phi_i(\mathbf{x}),
\end{align}
where $w_i$ is the weight assigned to the $i^{\rm th}$ specification and $\phi_i(\mathbf{x})$ is its normalized penalty term.
For one-sided specifications, the penalty term is defined as:
\begin{align}
    \phi_i(\mathbf{x}) &= \max\left(0, 1 -\left(\frac{g_i(\mathbf{x})}{S_i}\right)^{\sigma_i} \right),
    \label{eq:inequality}
\end{align}
where, $\sigma_i = +1$ for a constraint $g_i(\mathbf{x}) \geq S_i$, and $\sigma_i = -1$ for a constraint of the type $g_i(\mathbf{x}) \leq S_i$.  Specifications that must match a desired value are modeled by the normalized error:
\begin{align}
    \phi_i(\mathbf{x}) =
    \frac{|g_i(\mathbf{x}) - S_i|}{|S_i|}.
    \label{eq:equality}
\end{align}

Eqs.~\eqref{eq:inequality} and \eqref{eq:equality} ensure that there is no penalty if the specification is met, and the penalty increases as the magnitude of the violation increases. Hence, $C_{DE}(\mathbf{x})=0$ indicates full satisfaction of all target specifications. 

The outputs of this step correspond to the \emph{optimized device sizes} block in Figure~\ref{fig:flowchart}, and include the required total fin count $F_k^{\mathrm{req}}$ for each active primitive and required passive-primitive sizes. These are passed to the subsequent primitive-realization stage, along with other optimized variables such as biasing sources that will be used in the final simulation.

\noindent
\textbf{Step 2: Quantized-Height Primitive Generation.}
The optimized device sizes are next translated into parameterized analog primitives under fixed-height constraints. Since a given circuit function may admit multiple discrete FinFET realizations, this stage generates and evaluates candidate primitive implementations that differ in quantities such as finger partitioning, multiplicity, and stack configuration. The result is a set of height-compatible primitive candidates suitable for structured assembly. The primitive-generation and selection procedure is described in detail in Section~\ref{sec:layout_synthesis}.

\noindent
\textbf{Step 3: Row-Based Quantized-Height Placement and Routing.}
The selected primitive set is then passed to the physical synthesis stage, where the primitives are assembled in a row-based layout fabric and connected through constrained routing. At this stage, the layout must preserve the regularity implied by the quantized-height abstraction while satisfying analog-aware requirements such as device-type row organization, symmetry, proximity, and routing feasibility. The detailed PnR methodology is presented in Section~\ref{subsec:placement_n_routing}.

\noindent
\textbf{Step 4: Post-Layout Verification.}
After placement and routing, the generated layout is extracted to produce a parasitic-annotated netlist for simulation. The resulting performance metrics are then returned with the final GDSII layout.

%% file: contents/4_Row_based_layout_synthesis.tex
\section{Quantized-height layout synthesis}
\label{sec:layout_synthesis}

\subsection{Fixed height analog primitive generation}
\label{subsec:primitive_generation}
\noindent 
Our proposed methodology is built on \emph{analog primitives}~\cite{Kunal19, Dhar21}, which serve as the smallest parameterizable active and passive analog building blocks in the layout synthesis flow. These primitives are instantiated from a predefined parameterized library whose layout is synthesized using script-driven generators that produce design-rule-check (DRC)-compliant layout templates in a consistent, gridded style.

\begin{figure}[t]
    \centering
\includegraphics[width=1\linewidth]{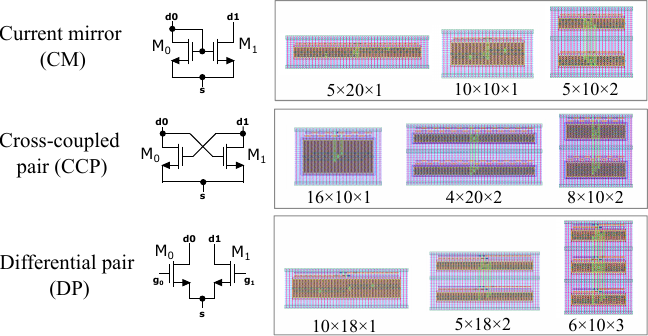}
    \caption{Examples of parameterizable active primitives and their corresponding layouts under different discrete sizing tuples, expressed as $(\textit{fin} \times n_f \times m)$.}
    \label{fig:primitives}
\end{figure}

\begin{figure}[t]
    \centering    
\includegraphics[width=0.95\linewidth]{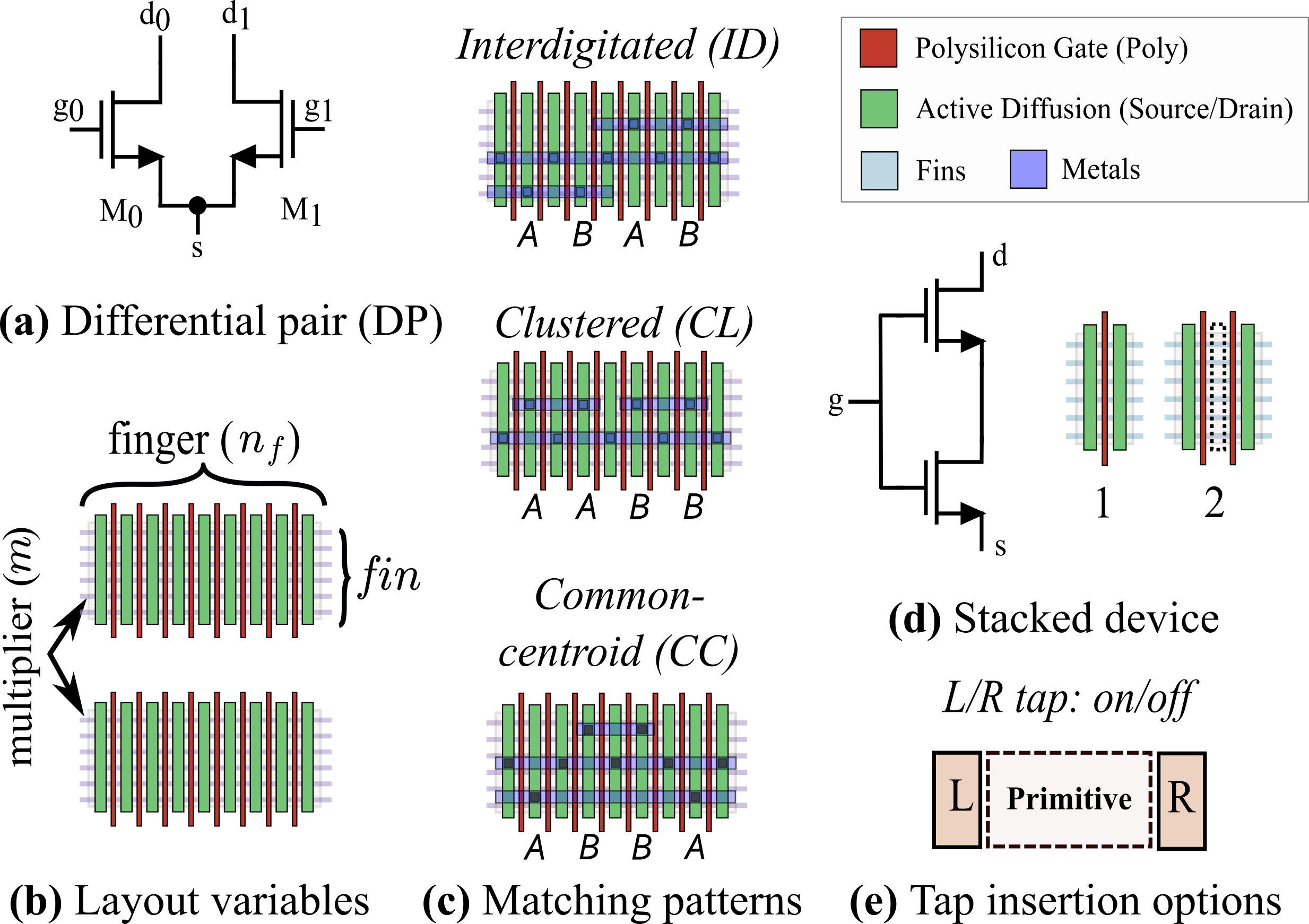}
    \caption{Primitive-cell layout-level design variables and implementation options: (a)~DP primitive schematic; (b)~layout notation for $fin$, $n_f$, and $m$; (c)~matching patterns within a shared diffusion region; (d)~stacked-device layout; and (e)~tap-location options for tap insertion.}
    \label{fig:primitives_dof}
\end{figure}

Representative primitives composed of active devices are shown in Figure~\ref{fig:primitives}. Each primitive implements a specific analog function using a small set of device units, which adhere to schematic-derived quantities such as the \textit{device width} and \textit{device matching patterns} propagated from the parent netlist. Multiple variants of a primitive layout may be built, parameterized by degrees of freedom such as the number of fins (\textit{fin}), fingers ($n_f$), and multipliers ($m$), illustrated in Figure~\ref{fig:primitives_dof}(b) for the DP primitive shown in Figure~\ref{fig:primitives_dof}(a). The layouts are generated in a manner that maximizes diffusion sharing.  Figure~\ref{fig:primitives_dof}(c) illustrates three device matching patterns: interdigitated (ID), clustered (CL), and common-centroid (CC). Our approach generates a set of possible layouts for each primitive, as shown in the figure. Layouts of various heights are generated, and a common optimal height for all cells is selected from among these layouts. The vertical spacing between neighboring device rows shown in Figure~\ref{fig:primitives} is selected to be DRC-compliant while providing sufficient routing and power delivery network (PDN) resources, including support for local gate/poly connections and lower-metal power rails. Unlike digital standard-cell layout, which embeds both \textit{VDD} and \textit{GND}  rails at predefined vertical locations, our primitives may be NMOS-only or PMOS-only and may not be connected to both rails (e.g., CCP (M$_5$,M$_6$) in Figure~\ref{fig:fixed-height-arch}; in some cases, the cell may be connected to no rail (e.g., DP(M$_1$,M$_2$)). Therefore, in our methodology, supply connectivity is resolved during routing against the global power grid.

To remain compatible with the discretized layout fabric of advanced-node technologies, we enforce strict poly- and fin-grid alignment, such that effective channel lengths are realized through series stacking of minimum-length devices, as shown in Figure~\ref{fig:primitives_dof}(d). This choice preserves pitch-constrained layout regularity while enabling control of analog operating points (e.g., output resistance) and improving matching robustness under these discretization constraints. Finally, a primitive layout may have a well tap associated with it, placed on its left (L) or the right (R) or both, as shown in Figure~\ref{fig:primitives_dof}(e).

As mentioned earlier, to ensure placement regularity and composability, the primitives adopt a quantized-height abstraction where each cell height in the primitive is expressed as an integer multiple of a base row height (e.g., $1\times, 2 \times$), analogous to digital standard cells, while retaining more flexible height selection than a fixed set of track-based cell offerings. This is achieved by varying $m$ while keeping the total $\textit{fin}$ fixed, e.g., in the $5 \times 20 \times 1$ and $5 \times 10 \times 2$ CM layouts in Figure~\ref{fig:primitives}. This structured primitive abstraction follows the idea of encapsulating small groups of devices and local interconnect into reusable analog layout blocks~\cite{Kunal19}. In our approach, this abstraction is used to assemble the final layout as a set of height-compatible cells stacked like bricks in a wall. It provides three practical benefits for the quantized-height layout synthesis procedure:  (i)~encapsulation of intra-primitive devices and local interconnect reduces the burden on the top-level PnR engine, which is provided with optimized layouts for these commonly-used structures; (ii)~matching patterns and symmetry constraints are captured structurally within primitives; and (iii)~hierarchical composition at the primitive level enables scalable synthesis of larger analog blocks. Accordingly, analog primitives serve as the fundamental building blocks in our fixed-height analog layout synthesis framework.

\subsection{Parasitic-Aware ML-Driven Row Height Selection}
\label{subsec:height_selection_methodology}
\noindent
This section presents the proposed method for deriving the common row height, which is a key requirement for quantized-height layout synthesis. In the FinFET setting, this requirement is conveniently expressed through a common candidate fins-per-finger base, denoted by \textit{fin}$^{base}$. The selection of \textit{fin}$^{base}$ determines the height of the primitive cell and is guided by two complementary criteria: structural compatibility across all required active primitives and the electrical quality of the resulting realizations. To account for the latter efficiently, we employ an MLP-based surrogate model that predicts the electrical behavior of candidate realizations and incorporates it directly into the \textit{fin}$^{base}$ selection procedure in Algorithm~\ref{alg:fixed_height_selection}.

At the structural level, ideally, all active primitives in the circuit would admit an exact realization $(\textit{fin}_k,n_{f,k},m_k)$ with $\textit{fin}_k = \textit{fin}^{\textit{base}}$, so they can be instantiated as height-compatible cells with consistent abutment, pin alignment, and row regularity in the downstream layout flow. In practice, however, this requires all sizes chosen by the sizing algorithm to be integer multiples of $\textit{fin}^{\textit{base}}$, which is unlikely. Therefore, a single common \textit{fin}$^{base}$ rarely admits exact realizations for all primitives simultaneously: some primitives require the use of fin depopulation to adhere to the common height. This motivates the search procedure in Algorithm~\ref{alg:fixed_height_selection} (lines~\ref{line:init_exploration}--\ref{line:struct}).

\begin{figure}[t]
    \centering
    \includegraphics[width=1\linewidth]{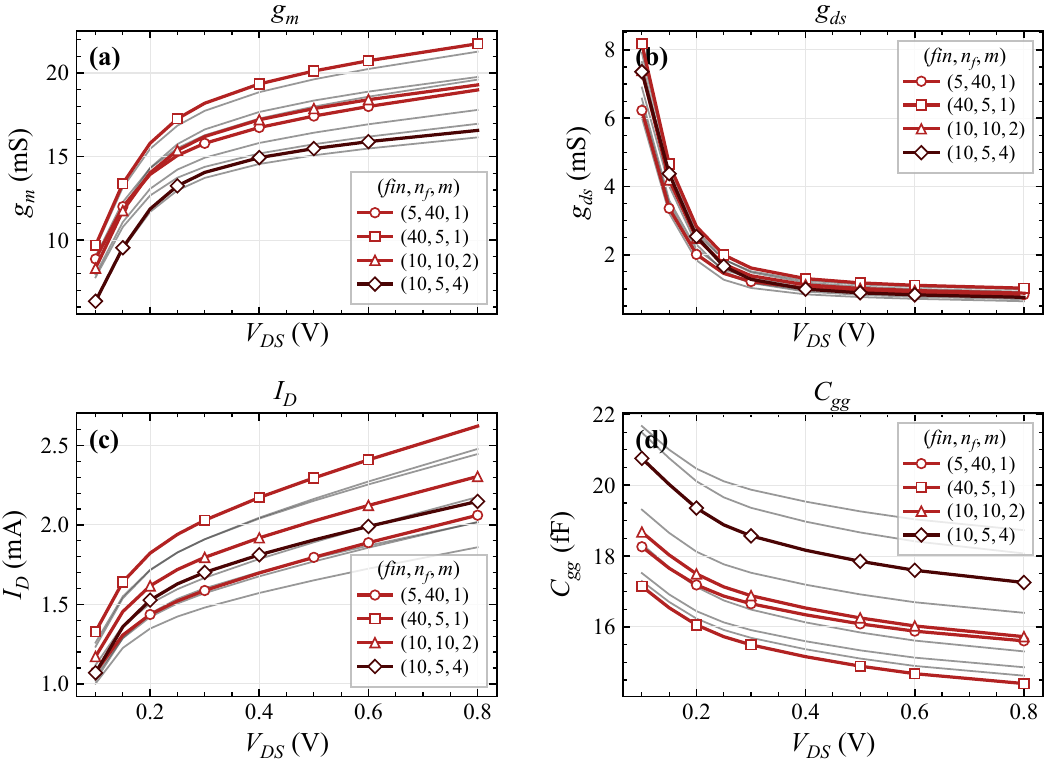}
    \caption{Parameter variation across device-realization tuples under a fixed total fin budget of \(\textit{Fin}^{\textit{req}}=200\): 
(a) $g_m$; (b) $g_{ds}$; (c) $I_D$; and (d) $C_{gg}$. Four example realizations are highlighted, while the remaining candidate realizations are shown in gray for visual clarity.}
    \label{fig:electrical_characteristics}
\end{figure}

When exact matching is not possible, a primitive is approximated by a nearby realizable tuple (line~\ref{line:approximation}). Specifically, the approximation selects the nearest realizable $(\textit{fin}_k,n_{f,k},m_k)$ combination 
% under the candidate $\textit{fin}^{\textit{base}}$ 
by allowing bounded fin depopulation such that
\begin{equation}
\textit{fin}_k \leq \textit{fin}^{\textit{base}}
\end{equation}
to approximate the required number of fins, $\textit{Fin}_k^{req}$. This approximation introduces a structural mismatch between the required and realized total fin counts, which we penalize using the normalized structural cost (line~\ref{line:struct}) to quantify the deviation from the desired $\textit{Fin}_k^{req}$. Accordingly, the first objective of the \textit{fin}$^{base}$ selection stage is to identify bases that preserve fixed-height compatibility while minimizing the structural error introduced by approximation.

\input{algorithms/algorithm_1_fixed_height_selection}

Structural feasibility alone, however, is insufficient. Different realizable tuples that satisfy the same common-base constraint can still produce different parasitic and electrical behavior. Changing the realized tuple $(\textit{fin}_k,n_{f,k},m_k)$ alters device folding, gate resistance, parasitic loading, and interconnect structure, and therefore changes the primitive's electrical behavior even when the nominal total fin count remains similar. Figure~\ref{fig:electrical_characteristics} illustrates this effect: realizations having the same total fin budget can exhibit noticeably different electrical response across both bias conditions and tuple choices. In the figure, a device with $\textit{Fin}^{\textit{req}}=200$ is evaluated across multiple realization tuples. At $V_{DS}=V_{GS}=0.4\,\mathrm{V}$, the maximum spread across realizations reaches $33.7\%$ in $g_m$ and $42.4\%$ in $I_D$ relative to their mean values. This observation motivates the need for a parasitic-aware electrical evaluation alongside the structural feasibility test.

To incorporate electrical behavior into the \textit{fin}$^{base}$ selection procedure, selected device electrical parameters, namely transconductance ($g_m$), output conductance ($g_{ds}$), drain current ($I_D$), and total gate capacitance ($C_{gg}$), are combined to form an electrical scoring model. These parameters are obtained from a modified netlist representation in which the base device response is augmented with estimated interconnect overhead associated with the chosen folding and multiplier configuration. To avoid repeated SPICE simulation of candidate primitive realizations during the search, the algorithm evaluates each feasible tuple using an MLP-based surrogate model (line~\ref{line:predictor}). We use two trained surrogate models, one for P-type devices and one for N-type devices. Each surrogate is implemented as a fully connected regressor with three hidden layers of sizes $10$, $20$, and $20$, with ReLU activations and a log-standardization transform applied to the targets.

\begin{figure}[t]
    \centering
    \includegraphics[width=1\linewidth]{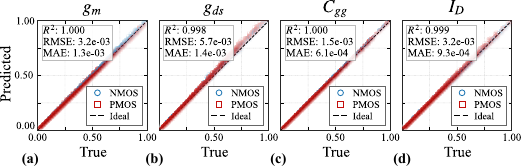}
    \caption{Normalized true-versus-predicted MLP accuracy plots for N- and P-type device electrical parameters: (a) $g_m$; (b) $g_{ds}$; (c) $C_{gg}$; and (d) $I_D$. Reported metrics are computed over the combined test set.}
    \label{fig:ml_accuracy}
\end{figure}

The surrogate models act as parasitic-aware electrical estimators for the quantities $(\hat g_{m,k}, \hat g_{ds,k}, \hat I_{D,k}, \hat C_{gg,k})$ from the realized tuple and operating point,
\begin{equation}
 (\text{type}_k,\; \textit{fin}_k,\; n_{f,k},\; m_k,\; V_{GS,k},\; V_{DS,k}),   
\end{equation}
where $\text{type}_k$ is the device type and $(V_{GS,k},V_{DS,k})$ are obtained from the preceding sizing stage. The accuracy of this surrogate is shown in Figure~\ref{fig:ml_accuracy}, which compares measured and predicted electrical parameters for N- and P-type devices.

From these predicted quantities, candidate primitive realizations are scored using device-level metrics related to speed, gain, and efficiency (line~\ref{line:score}). Specifically, we use a gate-resistance-aware speed metric derived from the transit-frequency relation, $m_{\text{spd}}$, with intrinsic-gain and transconductance-efficiency metrics, denoted by $m_{\text{gn}}$ and $m_{\text{eff}}$, respectively\cite{Carusone12, Jespers17}:
\begin{equation}
\{m_{\text{spd}}, m_{\text{gn}}, m_{\text{eff}}\} =
\left\{
\frac{\hat g_{m,k}}{\hat C_{gg,k}\bigl(1+\hat g_{m,k}R_{g,k}\bigr)},
\frac{\hat g_{m,k}}{\hat g_{ds,k}},
\frac{\hat g_{m,k}}{\hat I_{D,k}}
\right\}.
\end{equation} 
Here, the gate resistance is modeled using the distributed gate-resistance model in \cite{razavi94, saad22} as
$R_{g,k} = R_{\mathrm{sh},g}W_{f,k}/(3\,n_{f,k}L_k)$;

where $R_{sh, g}$ is the gate sheet resistance, $L_k$ is the channel length, while $W_{f, k}$ is the gate width per-finger.

\begin{figure}[t]
    \centering
    \includegraphics[width=1\linewidth]{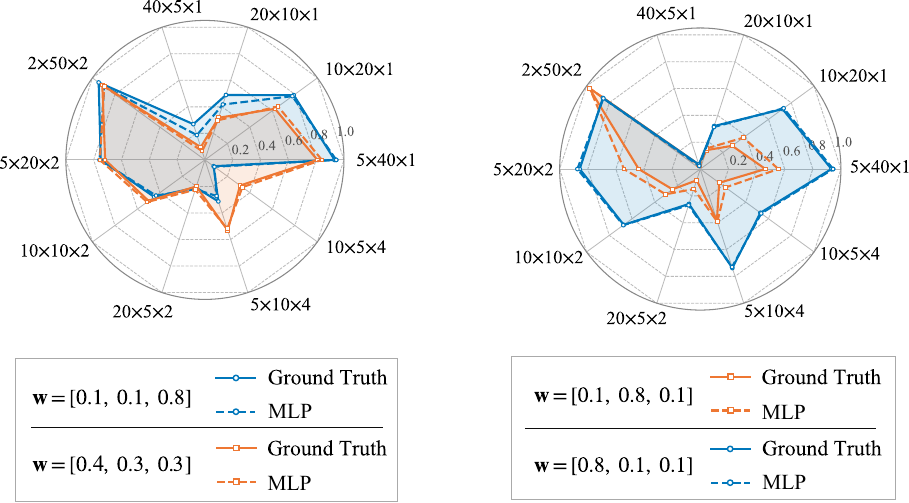}
    \caption{Electrical score comparison for different realization tuples $(\textit{fin} \times n_f \times m)$ under four different weight assignments using scores from ground-truth and MLP-predicted electrical parameters.}
    \label{fig:score_comparison}
\end{figure}

These metrics are then combined into a weighted score,
\begin{equation}
\mathcal{S}_k =
\mathbf{w}^{\top}
\begin{bmatrix}
\tilde m_{\mathrm{spd},k} &
\tilde m_{\mathrm{gn},k} &
\tilde m_{\mathrm{eff},k}
\end{bmatrix}^{\top},
\label{eq:electrical_score_final_short}
\end{equation}
where $\mathbf{w}=[w_0,w_1,w_2]$ is the role-weight vector, with $0<w_i<1$ for $i\in\{0,1,2\}$ and $\sum_{i=0}^{2} w_i=1$. Here, $w_0$, $w_1$, and $w_2$ represent the relative emphasis on speed, gain, and efficiency, respectively. Accordingly, bandwidth-critical, gain-critical, and efficiency-critical primitives can be ranked differently under the same framework. Figure~\ref{fig:score_comparison} further illustrates that different role-weight settings favor different primitive realizations. For example, with $\mathbf{w}=[0.1,\,0.1,\,0.8]$, the $(2\times50\times2)$ realization becomes preferable due to the stronger emphasis on transconductance efficiency, while with $\mathbf{w}=[0.8,\,0.1,\,0.1]$, the realization $(5\times40\times1)$ achieves the highest score, indicating that it is preferred when the emphasis is placed mainly on speed. Under the more balanced setting $\mathbf{w}=[0.4,\,0.3,\,0.3]$, the larger-finger realization $(2\times50\times2)$ still shows a slight advantage, mainly because its lower distributed gate resistance improves the speed-related score. The same realization is also favored for \(\mathbf{w}=[0.1,\,0.8,\,0.1]\), where its lower \(g_{ds}\) improves the intrinsic-gain-related score. The figure also confirms that this ranking trend is preserved when the score is computed from MLP-predicted quantities, as indicated by the broken lines.

Finally, Algorithm~\ref{alg:fixed_height_selection} aggregates the structural mismatch and the electrical score for each candidate common base, normalizes the candidate-level quantities, and combines them to select the best structural--electrical trade-off (lines~\ref{line:aggregate}--\ref{line:select}). The winning base $\textit{fin}^{\textit{base}*}$ and its associated $\{(\textit{fin}^*_k,n_{f,k}^*,m_k^*)\}_{k=1}^{K}$ are then passed to the fixed-height primitive generator. The resulting score therefore acts as a physically informed preselection criterion before full layout-level validation.

\subsection{Row-Based Layout Assembly and Routing}
\label{subsec:placement_n_routing}
\noindent
After selecting the fixed primitive height using the procedure in Section~\ref{subsec:height_selection_methodology}, the flow proceeds to row-based layout assembly and routing.
We employ a staged MILP-based row placer with an A*-based router, implemented as a custom layout synthesis flow, to automate structured analog layout synthesis. The formulation supports geometric constraint injection (e.g., symmetry/proximity), well-aware placement, and optimized tap insertion.
Our flow adopts a structured abstraction that quantizes primitive geometry and routing interfaces. Each primitive occupies an integer multiple of a base row height, while lower metals (e.g., M1--M3) are reserved for intra-primitive routing, and pins are exposed on predefined routing tracks at an interface layer. This abstraction preserves regularity and cleanly interfaces the MILP placer with grid-based routing and early parasitic estimation.

\input{tables/table_2_placement_notation_summary.tex}

\noindent
\subsubsection{MILP-based well-aware row placement}
\label{sec:milp_placer}
\noindent MILP has been widely used in digital placement subproblems such as detailed placement and legalization by enabling exact handling of discrete placement decisions~\cite{Yan96, Li12}. The same optimization framework is also well suited to analog placement problems; in contrast to digital designs, analog blocks typically contain far fewer components and decision variables, which makes MILP formulation computationally practical while still allowing explicit enforcement of analog-specific constraints such as row quantization, symmetry, proximity, well clustering, and tap-aware refinement. Motivated by this, we decompose the proposed analog placer into three stages: 
\begin{itemize}[noitemsep,topsep=0pt,leftmargin=*]
\item Stage~1: row placement with well-aware attraction.
\item Stage~2: well-island extraction from Stage~1 placement. 
\item Stage~3: tap insertion, controlling horizontal displacement.
\end{itemize}

The resulting MILP sizes remain manageable for the analog-scale testcases targeted in this work. As a representative example, a 10-active-primitive analog circuit with two local well clusters produces a Stage~1 global placement MILP with $\sim$$300$ variables and $\sim$$600$ constraints, while Stage~3 forms the tap-aware refinement with $\sim$$2$K variables and $\sim$$6$K constraints. In this example, the complete two-MILP placement flow takes about $30$~s, with most of the runtime spent in Stage~1. For a 20-active-primitive circuit with the same two-cluster structure, Stage~3 increases to approximately $\sim$$7$K variables and $\sim$$20$K constraints. If partitioned into six local well clusters, the Stage~3 sizes reduce to approximately $\sim$$700$ variables and $\sim$$2$K constraints for 10 active primitives, going up to $\sim$$2.5$K and $\sim$$7$K, respectively, for 20 active primitives. Since our primitive-based layout abstraction pre-places groups of physical devices into composite blocks, it directly reduces the number of individual active and passive devices optimized during placement. Despite its larger formulation, Stage~3 is typically faster because the vertical placement and ordering decision are already fixed by Stage~1. The dominant scaling term in Stage~1 comes from the pairwise non-overlap constraint, leading to approximately quadratic growth with the number of primitive blocks. Stage~3 introduces additional tap-coverage variables, but the refinement is applied after the row assignment, vertical placement, and horizontal ordering are fixed, so it scales with the local refinement problem rather than reintroducing the full two-dimensional placement search. The dominant Stage~3 scaling term comes from the tap-coverage conductance model. Since each evaluation location may need to be checked against multiple candidate tap locations, the number of added variables grows approximately quadratically within each local well cluster, while well clustering and primitive grouping keep this growth localized; the detailed formulation is given in the next section. By separating global placement from fixed-row tap refinement, the proposed decomposition avoids a monolithic placement--tap MILP while remaining compatible with open-source solvers such as HiGHS/SciPy interface. For larger AMS systems, the same formulation can be applied hierarchically by refining local analog blocks first and then treating them as macros during higher-level assembly.

\noindent
\textbf{Stage~1: Well-aware row-based placement.} Stage~1 formulates the placement of fixed-height analog primitives as an MILP that simultaneously enforces legality and optimizes wirelength, encourages clustering of same-well active blocks, and overall layout compactness. In this stage, tap placement is not optimized and active primitives are assumed to be tapless. Let $\mathcal{C}$ denote the set of primitive blocks to be placed; each primitive block $i \in \mathcal{C}$ is characterized by the width, $w_i$, and a height that is an integer multiple of the row height $h$, i.e., $h_i = n_i h$, $n_i\in \mathbb{Z}^+$ (typically, $n_i \in [1,3]$). When allowed by the primitive definition, horizontal
flipping is modeled using a binary orientation variable $f_i\in\{0,1\}$, which selects between the original and mirrored primitive orientation.

The formulation jointly minimizes three objectives: a routing proxy based on half-perimeter wirelength (HPWL), a placement-span term for layout compactness, and a same-well attraction term that clusters active blocks (i.e., primitives with the same type of active device (NMOS or PMOS)) together, while enforcing non-overlap, die boundary constraints, row alignment, and spacing rules. User-specified analog structural constraints (e.g., symmetry and proximity) are incorporated and directly embedded into the MILP formulation: such constraints could also be inferred automatically~\cite{Kunal20,Kunal20ICCAD,Kunal23}.

\noindent
\textbf{Objective function.} Using the notation in Table~\ref{tab:milp_notation}, the row-based placement is formulated as:
\begin{equation}
    \textstyle \min \quad 
    \lambda_{\text{HPWL}}\mathrm{HPWL}_w
    + \lambda_{\text{well}} \sum_{(i,j)\in\mathcal{W}} d_{ij}
    + \lambda_{\text{area}} A
\label{eq:objective}
\end{equation}
The first term, the weighted HPWL, is defined as:
\begin{equation*}
\textstyle
\mathrm{HPWL}_w = \sum_{n \in \mathcal{N}} w_n 
\left[
(X_n^{\max} - X_n^{\min}) + (Y_n^{\max} - Y_n^{\min})
\right],
\end{equation*}
where $\lambda_{\text{HPWL}}$ is the global objective weight for wirelength and $w_n$ is a net-dependent relative weight to further tune the relative importance of nets in the placement. Auxiliary variables bound the coordinates
of all pins $p \in \mathcal{P}_n$ as
\begin{align}
X_n^{\min} \leq x_i + \delta_{p,x}(f_i), \quad &
X_n^{\max} \geq x_i + \delta_{p,x}(f_i),\\
Y_n^{\min} \leq y_i + \delta_{p,y}, \quad &
Y_n^{\max} \geq y_i + \delta_{p,y},
\end{align}
where $i$ is the index of the primitive containing pin $p$, and
$(x_i,y_i)$ denotes its bottom-left coordinates. The horizontal pin offset
$\delta_{p,x}(f_i)$ is orientation-dependent when horizontal flipping is
enabled; otherwise it is a fixed value, $\delta_{p,x}$.

Let $\mathcal{W}$ denote the set of unordered pairs of active blocks with the 
same well type. For each $(i,j)\in\mathcal{W}$, the attraction term penalizes the $L_1$ (Manhattan) distance
\begin{equation}
d_{ij} = |x_i-x_j| + |y_i-y_j|, 
% \label{eq:well_distance}
\end{equation}
These are easily converted to linear constraints. The third term promotes layout compactness by penalizing the span of the global placement bounding box. Instead of using the exact area, we use a weighted span surrogate
\begin{equation}
A = \rho \cdot \Delta X + (1-\rho) \cdot \Delta Y,
\label{eq:area_surrogate_old}
\end{equation}
where $\Delta X = X^{\max} - X^{\min}$ and $\Delta Y = Y^{\max} - Y^{\min}$, and $\rho = 1/(1+AR)$ controls the relative emphasis on horizontal versus vertical compactness based on the aspect ratio, $AR$. The global span variables are obtained as:
% linked to the placed block extents by
\begin{equation}
\begin{aligned}
X^{\min}\le x_i,\quad &
X^{\max}\ge x_i+w_i,\quad \\
Y^{\min}\le y_i,\quad &
Y^{\max}\ge y_i+h_i,
\qquad \forall i\in\mathcal{C}.
\label{eq:stage1_global_span}
\end{aligned}
\end{equation}

\noindent
\textbf{Boundary and row constraints.}
Each block $i$ is represented by its bottom-left corner $(x_i,y_i)$.
Boundary feasibility requires
\begin{equation}
    0 \le x_i \le W - w_i, \quad
    0 \le y_i \le H - h_i, \quad
    \forall i \in \mathcal{C}.
\end{equation}
Row alignment is enforced by requiring the bottom edge of each block to lie on the row grid. 
Let $r_i\in\mathbb{Z}$ denote the selected starting row of block $i$. 
Since block $i$, with height $h_i = n_i h$, spans \(n_i\) rows starting from row $r_i$,
\begin{equation}
    y_i = r_i h, \qquad
    r_i \in \{0,\dots,R-n_i\},
    \qquad \forall i\in\mathcal{C},
\label{eq:row_alignment}
\end{equation}
where $R=\lfloor H/h \rfloor$ is the total number of available row slots.

\noindent
\textbf{Non-overlap constraints.}
For every unordered pair of distinct primitive blocks $(i, j) \in \mathcal{C}$, at least one of the following four relative placements must hold: block $i$ is to the left of, right of, below, or on top of block $j$. Let $z_{ij}^{L},z_{ij}^{R},z_{ij}^{B},z_{ij}^{T}\in\{0,1\}$, respectively, be binary variables that activate these four cases, and let $M$ denote a sufficiently large constant used in the standard big-$M$ linearization. The resulting disjunctive non-overlap constraints, $\forall (i,j) \in \mathcal{C}$, where $i < j$, are:
\begin{align}
x_i + w_i + c_{ij}^{x} & \le x_j + M(1-z_{ij}^{L}), \\
x_j + w_j + c_{ij}^{x} & \le x_i + M(1-z_{ij}^{R}), \\
y_i + h_i + c_{ij}^{y} & \le y_j + M(1-z_{ij}^{B}), \\
y_j + h_j + c_{ij}^{y} & \le y_i + M(1-z_{ij}^{T}), \\
z_{ij}^{L}+z_{ij}^{R}+z_{ij}^{B}+z_{ij}^{T} & \ge 1.
\label{eq:nonoverlap}
\end{align}
Here, $c_{ij}^{x}$ is a pair-dependent minimum horizontal separation between blocks $i$ and $j$, derived from technology- and layout-rule-driven spacing requirements, which can be viewed as a rule-based horizontal keep-out margin between the occupied geometries of the two blocks. Similarly, $c_{ij}^{y}$ denotes the required vertical clearance. The condition $i<j$ ensures that each unordered block pair is enumerated just once.
% and avoids duplicate non-overlap constraints.

\noindent
\textbf{Stage~2: Graph-based well-island extraction.}
After Stage~1 placement, same-well active blocks are typically pulled closer, but may still remain fragmented due to spacing constraints and intervening blocks. Stage~2 extracts feasible well islands from the legalized placement using two-dimensional geometric checks as summarized in Algorithm~\ref{alg:well_island_clustering}. For each well type $\tau \in \{\mathrm{P},\mathrm{N}\}$, the procedure first filters the active blocks of the corresponding type to form the graph nodes $V_\tau$ (lines~\ref{line:well_type_loop}--\ref{line:well_filter_nodes}). The edge set $E_\tau$ is then initialized as empty before evaluating possible connections between blocks (line~\ref{line:well_init}). The procedure then builds an undirected graph whose nodes are the active blocks of type $\tau$, modeled as rectangles $(x_i,y_i,w_i,h_i)$. 
For each node $i\in V_\tau$, the algorithm queries only nearby same-well candidates $\mathcal{N}_i$ using the gap-threshold check $\textsc{GapWithin}$ (line~\ref{line:candidate}). This check keeps a candidate block $b_j$ only when the horizontal and vertical gaps between the two blocks are within the prescribed limits, i.e., $\Delta x(b_i,b_j)\le d_x^{\max}$ and $\Delta y(b_i,b_j)\le d_y^{\max}$. This candidate-query step also reduces runtime by avoiding direct enumeration of all $O(N^2)$ block pairs; instead, each block retrieves only nearby blocks whose bounding boxes lie in the specified geometric neighborhood. For each nearby candidate $j\in\mathcal{N}_i$, the algorithm evaluates whether the pair $(i,j)$ can belong to the same well island (line~\ref{line:well_candidate_loop}). It first uses $\textsc{AxisOverlap}$ to compute the overlap lengths $(\ell_x, \ell_y)$ along the $x$ and $y$ directions (line~\ref{line:well_overlap}). An edge $(i, j)$ is added when two additional conditions hold: (i) the blocks overlap by at least $\ell_{\min}$ along at least one axis, and (ii) the corridor between them is not obstructed by any other block (lines~\ref{line:well_overlap}--\ref{line:well_add_edge}). 
After all candidate pairs for the current well type have been examined, $\textsc{ConnComp}$ uses a depth-first search (DFS) traversal to extract the connected components of $(V_\tau, E_\tau)$, which define the well islands for type $\tau$ (line~\ref{line:well_cluster}).

\input{algorithms/algorithm_2_well_detection}

\noindent
\textbf{Stage~3: Placement MILP formulation with tap-insertion optimization.} Given the legalized row-based placement from Stage~1 and the well-island clusters extracted in Stage~2, Stage~3 performs a final local refinement that jointly optimizes tap insertion while limiting deviation from the Stage~1 placement. In this stage, the vertical coordinates and row assignments obtained in Stage~1 are kept fixed, while only the horizontal block positions are refined. This preserves the legality and analog structure established earlier, avoids re-solving the full placement problem, and reduces the optimization to a smaller tap-aware geometry refinement where taps are selectively inserted so that each well island receives sufficient tap coverage without unnecessary area overhead. 

\noindent
\textbf{Overview of the formulation.}
Let $\mathcal{T}\subseteq\mathcal{C}$ denote the set of tap-capable blocks,
i.e., active transistor-level primitives for which left/right well taps may be
inserted. Blocks outside $\mathcal{T}$, such as passive devices or higher-level
macros used in hierarchical placement, are treated as non-tap-capable. Let
$w_{\mathrm{tap}}$ denote the width of one optional tap region, assumed to be the same for all tap-capable primitives. Binary variables $t_i^L,t_i^R\in\{0,1\}$ indicate whether the left and right taps of block $i$ are inserted during Stage~3. Accordingly, the final occupied edges of a tap-enabled block are linear expressions of the tapless-body left-edge coordinate $x_i$. For $i\in\mathcal{T}$, the final occupied horizontal interval after tap insertion is
\begin{equation}
x_i^L = x_i - w_{\mathrm{tap}}t_i^L,
\qquad
x_i^R = x_i + w_i + w_{\mathrm{tap}}t_i^R .
\label{eq:stage3_extents}
\end{equation}
For non-tap-capable blocks \(j\notin\mathcal{T}\), this reduces to
\(x_j^L=x_j\) and \(x_j^R=x_j+w_j\). Therefore, for tap-capable blocks, the
occupied width is
\begin{equation}
x_i^R-x_i^L
=
w_i+w_{\mathrm{tap}}(t_i^L+t_i^R).
\label{eq:stage3_width}
\end{equation}

\noindent
\textbf{Objective function.}
Stage~3 minimizes tap usage while limiting deviation from the Stage~1 placement. Let $x_i^{(0)}$ be the Stage~1 tapless-body horizontal coordinate of block $i$, and $\Delta_i^x$ be a variable that bounds the absolute horizontal displacement: $\Delta_i^x \ge x_i-x_i^{(0)}, \Delta_i^x \ge x_i^{(0)}-x_i$. The Stage~3 objective is:
\begin{equation}
\begin{aligned}
\min \quad 
\textstyle
\lambda_{\text{tap}} \sum_{i \in \mathcal{T}} (t_i^L + t_i^R)
+ 
\lambda_{\mathrm{disp}}\sum_{i\in\mathcal{C}}\Delta_i^x .
\end{aligned}
\label{eq:stage3_objective}
\end{equation}
The first term penalizes unnecessary tap insertion, while the second term preserves the HPWL-aware placement structure obtained in Stage~1 by discouraging large horizontal movement during tap insertion.

\noindent
\textbf{Boundary and row-preserving non-overlap constraints}. The final horizontal span of each block including any inserted tap regions, must remain within the layout boundary:
\begin{equation}
0 \le x_i^{L}, 
\qquad
x_i^{R} \le W,
\qquad \forall i \in \mathcal{C}.
\label{eq:stage3_boundary}
\end{equation}
Since Stage~3 keeps the vertical coordinates fixed, vertical legality between two blocks cannot change during the refinement. Therefore, horizontal non-overlap need only be enforced for block pairs whose vertical spans have an overlap. We define the set of vertically conflicting pairs as
\begin{equation}
\begin{aligned}
\mathcal{O}=\{(i,j)\in\mathcal{C}\times\mathcal{C},\, i<j \mid
&\neg( y_i+h_i+c_{ij}^{y}\le y_j \\
&\vee\; y_j+h_j+c_{ij}^{y}\le y_i )\}.
\end{aligned}
\label{eq:stage3_overlap_pairs}
\end{equation}
For each pair \((i,j)\in\mathcal{O}\), let \(\ell(i,j)\) and \(r(i,j)\) denote the left and right blocks, respectively, according to the Stage~1 horizontal ordering. Stage~3 preserves this ordering by enforcing
\begin{equation}
x_{r(i,j)}^{L} \ge x_{\ell(i,j)}^{R} + c_{\ell(i,j),r(i,j)}^{x},
\qquad
\forall (i,j)\in\mathcal{O}.
\label{eq:stage3_nonoverlap}
\end{equation}
Here, \(c_{\ell(i,j),r(i,j)}^{x}\) is the same pair-dependent horizontal spacing model used in Stage~1. This order-preserving constraint avoids reintroducing the full four-way non-overlap disjunction in Stage~3 while still preventing overlaps among blocks that share the same vertical region.

\noindent
\textbf{Cluster-restricted tap coverage.}
We model tap coverage as a discrete approximation of the effective
well/substrate conductance from an evaluation point $q$ to nearby enabled taps.
Let $d_j(q)$ denote the distance from $q$ to enabled tap $j$. The resistance
from $q$ to tap $j$ is approximated as proportional to distance,
$R_j(q)=\kappa d_j(q)$, where $\kappa$ is a technology-dependent
resistance-per-unit-distance constant. The corresponding conductance is
therefore $1/(\kappa d_j(q))$, and multiple enabled taps contribute additively:
\begin{equation}
\textstyle 
G_{\mathrm{eff}}(q)
=
\sum_j 1/(\kappa d_j(q)).
\label{eq:tap_effective_conductance}
\end{equation}
Let $d_{\mathrm{tap}}$ denote the designer-specified effective tap distance
used to set the coverage threshold. Requiring the effective resistance to be no larger than
$R_{\mathrm{tap}}=\kappa d_{\mathrm{tap}}$ is equivalent to requiring
$G_{\mathrm{eff}}(q)\ge 1/R_{\mathrm{tap}}$, i.e.,
\begin{equation}
\textstyle
\sum_j 1/(d_j(q))
\ge
(1/d_{\mathrm{tap}}).
\label{eq:tap_conductance_requirement}
\end{equation}
Thus, nearby taps contribute more strongly than distant taps, and the minimum coverage requirement can be interpreted as a lower bound on the accumulated local tap conductance.  This requirement is evaluated at a finite set of deterministic coverage evaluation points. For each active block assigned to a well island, two evaluation points are created: one near the left side and one near the right side of the primitive bounding box. These points are placed along the vertical centerline of the block. For a block $i$, the left and right evaluation locations are placed at $x_i$ and $x_i + w_i$, respectively. 

\begin{figure}[t]
    \centering  \includegraphics[width=1\linewidth]{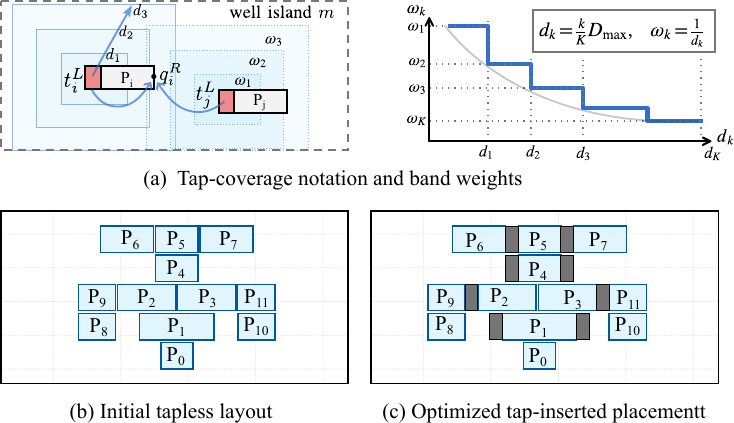}
    \caption{(a)~Tap-coverage notation and piecewise band-weight model; (b)~initial StrongARM comparator placement with all taps disabled; and (c)~optimized layout with selected taps enabled while preserving the Stage~1 placement structure and the required coverage constraint}
    \label{fig:tap_placement}
\end{figure}

The distance-dependent tap contribution is modeled by $K$ nested
rectangular coverage bands within a maximum candidate distance $D_{\max}$.
The $k^{\mathrm{th}}$ band boundary is defined as
\begin{equation}
    d_k = (k/K) \cdot D_{\max}, \qquad k=1,\ldots,K .
\end{equation}
Following the inverse-distance conductance model, the weight assigned to band
$k$ is chosen using the corresponding band boundary, $\omega_k = 1/d_k, k=1,\ldots,K$ as shown in Figure~\ref{fig:tap_placement}(a). Let $\mathcal{Q}_m$ denote the set of coverage evaluation points in well
island $m$. For active blocks with tap side $s \in \{L, R\}$, evaluation point $q \in \mathcal Q_m$, and band index $k$, the binary variable $v^{(s)}_{pqk}$ indicates that $q$ lies within the $k^{th}$ coverage band of tap side $s$ of block $p$, and that the corresponding tap side is enabled.
The geometric implication linking $v_{pqk}^{(s)}$ to the coverage band is enforced with big-M constraints:
\begin{equation}
    |u_q-u_p^{(s)}| \le d_k+M(1-v_{pqk}^{(s)}), \qquad u \in \{x, y\}.
\end{equation}
The band variables are nested and active only when the corresponding tap side is enabled:
\begin{align}
v_{pqk}^{(s)} & \le v_{pq,k+1}^{(s)}, 
& \qquad k=1,\dots,K-1,
\label{eq:stage3_nested_bands}
\\
v_{pqk}^{(s)} & \le \tau_p^{(s)},
& \qquad k=1,\dots,K,
\label{eq:stage3_tap_enable_link}
\end{align}
where $\tau_p^{(s)}$ denotes the side-dependent tap-enable variable, corresponding to $t_p^L$ for $s=L$ and $t_p^R$ for $s=R$. The contribution of tap side $s$ is
\begin{equation}
\textstyle
c_{pq}^{(s)} = \omega_K \cdot v_{pqK}^{(s)} +
\sum_{k=1}^{K-1} \left(\omega_k - \omega_{k+1}\right) \cdot v_{pqk}^{(s)} .
\label{eq:stage3_weighted_score}
\end{equation}
Nested band variables are used because the rectangular coverage regions are cumulative: if an evaluation point lies in a closer band, it also lies in every larger band. The nesting constraints therefore match the geometry directly and the encoding recovers the intended band weight. An alternative would be to use one-hot encoding, which requires redefining the bands as exclusive rectangular rings. For each ring the point must be inside the outer rectangle and outside the previous inner rectangle, but the latter condition is disjunctive for rectangular regions and requires additional big-M logic. We therefore use the nested-band encoding.

The net tap-conductance cost at evaluation point $q \in \mathcal{Q}_m$ is
\begin{equation}
\textstyle
I_q = \sum_{p\in\mathcal{T}_m}
\left(
c_{pq}^{(L)} + c_{pq}^{(R)}
\right),
\end{equation}
Here, $\mathcal{T}_m=\mathcal{T}\cap\mathcal{G}_m$ denotes the tap-capable blocks within well island
$m$, where
$\mathcal{G}_m$ is the set of blocks in that well island. Terms whose maximum coverage rectangle cannot contain $q$ are omitted. The total tap requirement is enforced as:
\begin{equation}
I_q \ge I_{\min},
\qquad \forall q\in\mathcal{Q}.
\label{eq:stage3_intensity_lb}
\end{equation}
where $I_{\min}$ corresponds to the target inverse-distance threshold
$1/d_{\mathrm{tap}}$. This parameter gives the designer a direct knob for controlling tap density: larger $I_{\min}$ values require stronger local tap conductance and therefore denser tap insertion, while smaller values permit sparser tap placement. This is useful in analog layouts because tap requirements are not only a matter of satisfying the maximum tap spacing in the process design rule manual; sensitive analog circuits may require more frequent taps to better control well/substrate potential~\cite{Williams23}.

Figure~\ref{fig:tap_placement}(a) illustrates the notation used in the cluster-restricted tap-coverage model. For each well island, evaluation points are placed near the left and right edges of active primitives, and only taps belonging to the same well island contribute to the corresponding coverage sum. Enabled tap sides contribute through nested rectangular bands with piecewise-constant weights derived from the inverse-distance proxy $\omega_k=1/d_k$. The binary band variables therefore couple tap enablement, band membership, and the accumulated coverage score $I_q$. The Stage~3 MILP optimizes insertion of taps while enforcing $I_q\ge I_{\min}$ at all evaluation points. Figure~\ref{fig:tap_placement}(b)--(c) shows the transition from the tapless Stage~1 StrongARM comparator placement to the Stage~3 tap-inserted layout, where only the tap regions needed to satisfy the local coverage requirement are inserted while limiting horizontal displacement.

\noindent
\textbf{Structural constraints.}
All user-defined analog structural constraints introduced in Stage~1 (e.g., symmetry and proximity) are preserved in Stage~3 by reapplying the same external constraint interface to the reduced one-dimensional horizontal refinement problem. Tap-specific structural constraints are also supported by directly coupling tap-enable binaries, e.g., $t_i^L=t_i^R$ for self-symmetric tap insertion, or $t_i^L=t_j^R$ and $t_i^R=t_j^L$ for a symmetric pair $(i,j)$. For example, the StrongARM comparator placement in Figure~\ref{fig:tap_placement}(b) and (c) is structurally symmetric because the user-injected symmetry constraint is retained during the placement stages.

\subsubsection{Track-Constrained Row-Based Routing}
\begin{figure}[t]
    \centering
\includegraphics[width=1\linewidth]{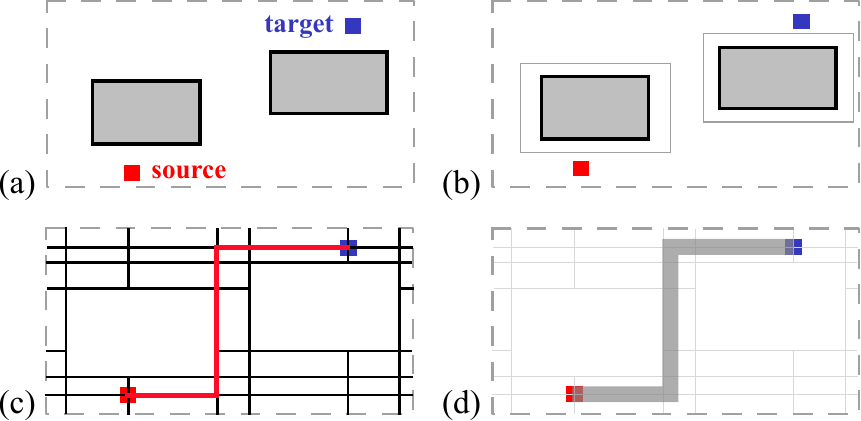}
\caption{Routing: (a)~source, target pins and existing metal shapes; (b)~obstacles are \textit{bloated} by (spacing+width/2); (c)~routing grid using source, target, and bloated obstacles, and the A$^*$ solution; and (d)~solution shapes.}
\label{fig:hanan_grid}   
\end{figure}

The proposed framework employs a grid-based, detailed router suitable for row-based fixed-height analog layouts in FinFET technologies. Unlike unconstrained polygonal routing, the router operates under track and direction restrictions imposed by advanced-node back-end stacks. Each routing layer is assigned its preferred direction, while vias provide transitions between adjacent layers. A discrete routing graph is constructed from the existing metal geometry in each layer. As illustrated in Figure~\ref{fig:hanan_grid}(a), the input consists of source and target pins together with existing metal shapes arising from placed primitives, routing blockages, or previously routed nets. To ensure DRC-compliant routing, each existing shape is converted into an obstacle by bloating its boundary by the minimum spacing rule plus half of the intended routing width, as shown in Figure~\ref{fig:hanan_grid}(b). The routing grid is then formed by introducing horizontal and vertical grid lines at the $x$- and $y$-coordinates of source and target pins as well as obstacle boundaries~\cite{Hanan66}. Grid segments lying fully inside obstacles are removed, yielding the reduced routing graph in Figure~\ref{fig:hanan_grid}(c).

Let $G=(V,E)$ denote the resulting routing graph, where each node $v\in V$ corresponds to a legal grid location on a routing layer, and each edge $e\in E$ represents either an intralayer wire segment or an interlayer via transition. Routing between a source $s$ and a target $t$ is the shortest-path problem:
\begin{equation}
\textstyle
P^*=\arg\min_{P:s\rightarrow t}\sum_{e\in P} c(e),
\label{eq:routing_path}
\end{equation}
where the edge cost is defined as
\begin{equation}
c(e)=
\begin{cases}
\alpha_\ell\,L(e), & e \text{ is a wire segment on layer } \ell,\\[3pt]
\beta_v, & e \text{ is a via transition,}
\end{cases}
\label{eq:routing_cost}
\end{equation}
with $L(e)$ the segment length, $\alpha_\ell$ the layer-dependent routing cost, and $\beta_v $ the via penalty. For multi-terminal nets, the router greedily builds a connection tree by first selecting the closest pin pair and then repeatedly connecting the nearest unconnected pin to the partially routed net. For each selected source--target connection, A* search is applied on $G$ to obtain the minimum-cost path $P^*$~\cite{Hart68}. Thus, the selected route is not merely the geometrically shortest path, but the minimum-cost path under layer-dependent wire and via penalties. The final path is converted into metal geometry by expanding the centerline by half of the routing width in the permitted direction, as illustrated in Figure~\ref{fig:hanan_grid}(d).
Although Fig.~\ref{fig:hanan_grid} shows a two-dimensional slice, the implemented router operates on a three-dimensional multilayer graph. Obstacles are generated independently on each layer, aligned grid locations are formed across the stack, vias connect corresponding grid points on adjacent layers, and segments violating blockages/directional constraints are removed. Since routing order affects later pin accessibility, the framework explores multiple net-order permutations when needed; in practice, runtime remains on the order of seconds for the target testcases.

%% file: algorithms/algorithm_1_fixed_height_selection.tex
\begin{algorithm}[t]
\caption{Parasitic-aware common fin exploration for quantized-height primitives}
\label{alg:fixed_height_selection}
\footnotesize
\begin{algorithmic}[1]
\Require $\{\textit{Fin}_k^{\textit{req}}\}_{k=1}^{K}$ (required total fins) for $K$ primitives; 
\Statex $\mathcal{F}$ (candidate common bases); 
\Statex $N_f^{\min},N_f^{\max}$ (finger-count bounds); 
\Statex $\textit{fin}^{\min}$ (minimum fins per finger); 
\Statex $AR_{\max}$ (aspect-ratio limit); 
\Statex $d_{\max}$ (maximum depopulation); 
\Statex $\{(V_{GS,k},V_{DS,k})\}_{k=1}^{K}$ (operating points); 
\Statex $\mathcal{N}(\cdot)$ (MLP predictor); 
\Statex $\mathcal{S}(\cdot)$ (role-dependent score model); 
\Statex $\hat{\mathbf{y}}_k = (\hat g_{m,k},\hat g_{ds,k},\hat I_{D,k},\hat C_{gg,k})$ (electrical parameters);  
% \Statex $k$ indexes of the primitive; 
\Statex $\{\mathrm{role}_k\}_{k=1}^{K}$ (primitive role-weight)
\Ensure $\textit{fin}^{\textit{}{base}*}$ (best common base) 
\Statex $\{(fin_k,n_{f,k},m_k)\}_{k=1}^{K}$ (assigned realizations)
\Statex \textbf{Step 1: Evaluate each candidate common base}
\Statex \LeftComment{Iterate over the candidate common-base fin values}
\ForAll{$\textit{fin}^{\textit{base}} \in \mathcal{F}$} 
    \For{$k=1$ to $K$} \label{line:init_exploration} \LeftComment{Iterate over the active primitives}
    % \red{Ditto. And is $K$ ever defined?}
        \Statex \hspace*{8mm} \LeftComment{Search for feasible $(\textit{fin}_k, n_{f,k})$ that best matches $\textit{Fin}_k^{req}$}
        \State $m_k \gets$ smallest multiplier satisfying $AR_{\max}$ \label{line:multiplier}
        \State $(\textit{fin}_k,n_{f,k}) \gets \arg\min_{\textit{fin},n_f}|\textit{Fin}_k^{\textit{req}} - m_k n_f \textit{fin}|$ s.t. \label{line:search}
        \Statex \hspace{16mm} $n_f \in [N_f^{\min},N_f^{\max}]$, 
        \Statex \hspace{16mm} $\textit{fin} \in [\textit{fin}^{\textit{base}}-d_{\max},\textit{fin}^{\textit{base}}]$
        \Statex \hspace{16mm} $\textit{fin} \ge \textit{fin}^{\min}$ \label{line:approximation}
        \Statex \hspace{8mm} \LeftComment{Add structural penalty} 
        \State $\epsilon_k \gets |\textit{Fin}_k^{\textit{req}} - m_k n_{f,k} \textit{fin}_k| / \textit{Fin}_k^{\textit{req}}$ \label{line:struct}
        \State $\hat{\mathbf{y}}_k \gets \mathcal{N}(\text{type}_k,\textit{fin}_k,n_{f,k},m_k,V_{GS,k},V_{DS,k})$ \label{line:predictor}
        \Statex \hspace{8mm} \LeftComment{Determine electrical score} 
        \State $\sigma_k \gets \mathcal{S}(\hat{\mathbf{y}}_k;\mathrm{role}_k)$ \label{line:score}
    \EndFor
    \State $\bar C(\textit{fin}^{\textit{base}}) \gets \frac{1}{K}\sum_{k=1}^{K}\epsilon_k$; \quad $\bar S(\textit{fin}^{\textit{base}}) \gets \frac{1}{K}\sum_{k=1}^{K}\sigma_k$ \label{line:aggregate}
\EndFor
\Statex \textbf{Step 2: Normalize candidate costs and scores}
% \Statex \hspace*{-0.5em}\Comment{Perform min--max normalization of $\bar C$ and $\bar S$ over all candidate bases in $\mathcal{F}$}
\State $\widetilde{C},\widetilde{S} \gets \text{Normalize }\bar C,\bar S \text{ over }\mathcal{F}$ \label{line:norm}
\State $\textit{fin}^{\textit{base}*}=\arg\min_{\textit{fin}^{\textit{base}}}\left[\lambda_s \widetilde{C}(\textit{fin}^{\textit{base}})+\lambda_e(1-\widetilde{S}(\textit{fin}^{\textit{base}}))\right]$ \label{line:select}

\Statex \textbf{Step 3: Select the best common base}
\State \Return $\textit{fin}^{\textit{base*}}$ and the associated $\{(\textit{fin}_k,n_{f,k},m_k)\}_{k=1}^{K}$
\end{algorithmic}
\end{algorithm}

%% file: tables/table_2_placement_notation_summary.tex
\begin{table*}[t]
\centering
\caption{Notation summary for the three-stage placement formulation.}
\label{tab:milp_notation}
\scriptsize
\setlength{\tabcolsep}{3pt}
\renewcommand{\arraystretch}{1.03}
\begin{tabularx}{\textwidth}{@{} l X l X @{}}
\toprule
\textbf{Symbol} & \textbf{Description} & \textbf{Symbol} & \textbf{Description} \\
\midrule

$\mathcal{C},w_i,h_i,h,n_i$
& Set of primitive blocks, tapless primitive width, height of block \(i\), base row height, row-height multiplier, with \(h_i=n_i h\)
&
$d_x^{\max},d_y^{\max},\ell_{\min},\mathcal{G}_m$
& Stage~2 horizontal/vertical gap thresholds, minimum overlap threshold, extracted well-island clusters \(\mathcal{G}_m\) \\

$\lambda_{\mathrm{HPWL}},\lambda_{\mathrm{well}},\lambda_{\mathrm{area}},\mathcal{W},d_{ij},A$
& Stage~1 objective weights, set of same-well block pairs, same-well distance for pair \((i,j)\in\mathcal{W}\), area surrogate
&
$\mathcal{T},w_{\mathrm{tap}}, d_{\mathrm{tap}}$
& Set of tap-capable blocks, tap region width, designer-specified effective tap distance
\\

$\mathcal{N},w_n,X_n^{\min},X_n^{\max},Y_n^{\min},Y_n^{\max}$
& Set of nets, net-dependent HPWL weight, horizontal/vertical span variables for net $n$
&
$t_i^{L},t_i^{R},x_i^{L},x_i^{R}$
& Left/right tap-enable variables, final occupied horizontal extents of block $i$ \\

$\mathcal{P}_n,x_i,y_i,\delta_{p,x},\delta_{p,y},f_i$
& Set of pins for net $n$, bottom-left coordinates of block $i$, orientation-dependent $x$-offset \& fixed $y$-offset of pin $p$ from its parent block origin, 
% horizontal 
flip variable of block $i$
&
$\mathcal{Q}_m,\mathcal{Q},D_{\max},d_k,K$
& Island-specific/global coverage evaluation points, maximum candidate
coverage distance, band boundary, and number of coverage bands\\

$AR,X^{\min},X^{\max},Y^{\min},Y^{\max}$
& Aspect ratio and global placement-span limits
&
$\omega_k,v_{pqk}^{(s)},c_{pq}^{(s)},\tau_p^{(s)}$
& band weight, band-membership variable, tap-side contribution, tap-enable variable for side \(s\) of block $p$ \\

$W,H,r_i,R$
& Layout width, layout height, selected starting row of block $i$, total number of available row slots
&
$I_q,I_{\min}$ & Tap-conductance coverage proxy at evaluation point $q$, minimum required threshold\\

$z_{ij}^{L},z_{ij}^{R},z_{ij}^{B},z_{ij}^{T},M,c_{ij}^{x},c_{ij}^{y}$
& Non-overlap binary variables, big-$M$ relaxation constant, required horizontal/vertical clearance between blocks $i$ and $j$
&
$\Delta_i^x, \lambda_{\mathrm{tap}}, \lambda_{\mathrm{disp}}, \mathcal{O}$ & Stage~3 absolute horizontal displacement of block $i$, tap-count and displacement objective weights, set of vertically conflicting block pairs\\
\bottomrule
\end{tabularx}
\end{table*}

%% file: algorithms/algorithm_2_well_detection.tex
\begin{algorithm}[t]
\caption{Well-Island Clustering}
\label{alg:well_island_clustering}
\footnotesize
\begin{algorithmic}[1]
\Require Blocks $\mathcal{B}=\{b_1,\dots,b_N\}$, where
\Statex $b_i=(x_i,y_i,w_i,h_i,a_i,\tau_i)$;
\Statex $a_i\in\{\text{active},\text{passive}\}$ $\equiv$ block class,
$\tau_i\in\{\text{P},\text{N},\emptyset\}$ $\equiv$ well type;
\Statex thresholds $d_x^{\max}, d_y^{\max}$ (gap) and $\ell_{\min}$ (minimum overlap).
% \Statex Helpers: $\textsc{GapWithin}$, $\textsc{AxisOverlap}$, $\textsc{Blocked}$, $\textsc{ConnComp}$.%OK to skip

\Ensure Disjoint well-island clusters $\mathcal{G}_{\mathrm{P}},\mathcal{G}_{\mathrm{N}}$.
\Statex \LeftComment{Process one well type at a time}
\ForAll{$\tau \in \{\mathrm{P},\mathrm{N}\}$} \label{line:well_type_loop}
    \Statex \hspace*{4mm} \LeftComment{Filter nodes: Keep only active blocks of current well type} 
    \State $V_{\tau}\gets\{\,i \mid a_i=\text{active}\ \wedge\ \tau_i=\tau\,\}$ \label{line:well_filter_nodes}
    \Statex \hspace*{4mm} \LeftComment{Initialize}
    \State $E_{\tau}\gets\emptyset$ \label{line:well_init} 
    \Statex \hspace*{4mm} \LeftComment{Visit each node of the current well-type graph}
    \ForAll{$i \in V_{\tau}$} 
        \Statex \hspace*{8mm} \LeftComment{Candidate query: Collect nearby same-type candidates}
        \Statex \hspace*{8mm} \LeftComment{that satisfy the 2D gap thresholds} 
        \State $\mathcal{N}_i \gets \{\,j \in V_{\tau}\setminus\{i\}\mid \textsc{GapWithin}(b_i,b_j,d_x^{\max},d_y^{\max})\,\}$ \label{line:candidate}
        \ForAll{$j \in \mathcal{N}_i$} \label{line:well_candidate_loop}
            \Statex \hspace*{12mm} \LeftComment{Measure overlap of the two blocks along $x$, $y$ axes} 
            \State $(\ell_x,\ell_y)\gets \textsc{AxisOverlap}(b_i,b_j)$ \label{line:well_overlap}
            \Statex \hspace*{12mm} \LeftComment{If the pair has sufficient overlap and an unobstructed}
            \Statex \hspace*{12mm} \LeftComment{corridor, connect the two blocks in the graph}
            \If{$\max(\ell_x,\ell_y)\ge \ell_{\min}$ \textbf{and} $\neg\,\textsc{Blocked}(i,j,\mathcal{B})$}
                \State $E_{\tau}\gets E_{\tau}\cup\{(i,j)\}$ \label{line:well_add_edge}
            \EndIf
        \EndFor
    \EndFor
    \Statex \hspace*{4mm} \LeftComment{Cluster}
    \State $\mathcal{G}_{\tau}\gets \textsc{ConnComp}(V_{\tau},E_{\tau})$ \label{line:well_cluster}
\EndFor

\State \Return $\mathcal{G}_{\mathrm{P}},\mathcal{G}_{\mathrm{N}}$
\end{algorithmic}
\end{algorithm}

%% file: contents/5_Results.tex
\section{Experimental Results}
\label{sec:results}
\noindent
This section evaluates the proposed layout synthesis methodology and quantifies its impact on circuit performance, layout area, and PDN accessibility. All experiments are conducted with a 12nm FinFET technology using three families of analog and mixed-signal testcase circuits: OTAs, comparators, and filters. The testcase schematics and synthesized layouts are shown in Figure~\ref{fig:result_figure} and ~\ref{fig:result_figure_gmc}, and the corresponding physical and electrical metrics are presented in Tables~\ref{tab:benchmark_layout_summary}--\ref{tab:gmc_filter_performance}; for each circuit the complete flow is executed from schematic-level optimization through fixed-height primitive realization, placement, and routing to postlayout evaluation. 

We refer to layouts from our approach as quantized-height (QH) implementations. For comparison, we also implement a free-height (FH) baseline, where each analog primitive has an individually optimized height. Unlike the QH implementation, which constrains primitives to a row-based quantized-height fabric, the FH baseline allows a less constrained vertical arrangement of primitives and therefore serves as a proxy for conventional handcrafted analog layout styles. Based on this FH--QH comparison framework, we first report the schematic-level circuit optimization used to generate layout-compatible device targets. The layout-focused results are then presented in three parts: (i) a physical and electrical comparison between FH and QH layout implementations, (ii)  a row-height exploration study to quantify the performance tradeoffs across different quantized heights, and (iii) an evaluation of PDN accessibility and supply routing benefits of QH layout style.

\subsection{Circuit Design and Optimization}
\begin{figure}[t]
    \centering    \includegraphics[width=1\linewidth]{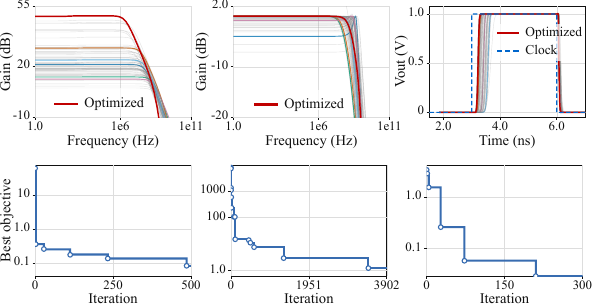}
   \caption{Representative optimization response waveforms and objective-function reduction for: (a)~folded-cascode OTA AC response; (b)~Gm-C filter AC response; and (c) StrongARM latch comparator transient response.} \label{fig:optimization}
\end{figure}

\noindent
Before layout synthesis, each testcase is optimized at the schematic level to obtain a feasible, layout-compatible design point. For the OTA-based blocks, including the standalone OTAs and the transconductance cells used inside the biquad Gm-C filter, the Phase~1 sizer initializes the design using a $g_m/I_D$-based sizing procedure. The resulting initial design is then refined in Phase~2 using the modular black-box optimization loop described in Section~\ref{sec:framework}, with  DE implemented through the \texttt{pymoo} framework~\cite{Blank20}. The StrongARM comparator bypasses the $g_m/I_D$ initializer and is optimized directly in Phase~2 using its transient performance objectives. The final schematic-level solutions provide the device targets used for fixed-height cell generation and subsequent layout synthesis. Figure~\ref{fig:optimization} summarizes representative optimization behavior for the main circuit classes used in this work. The response plots in the first row show the final optimized design waveforms together with a small subset of sampled candidate responses collected during the optimization process, with their convergence plots shown below the waveforms.

Across the evaluated testcases, ranging from the 5T~OTA to the Gm-C filter in circuit complexity, the schematic-level optimization, which provides an input to this work, involves $7$--$15$ circuit-level design variables, including device sizes and biasing sources, and requires ${\sim}2$--$9$~min, while the subsequent layout-generation stage, our main contribution, measured from ML-based fixed-height optimization to final GDSII generation, completes in ${\sim}20$--$90$~sec.

\subsection{Evaluation}
\subsubsection{Layout Comparison}
\begin{figure*}[t]
    \centering
\includegraphics[width=0.99\linewidth]{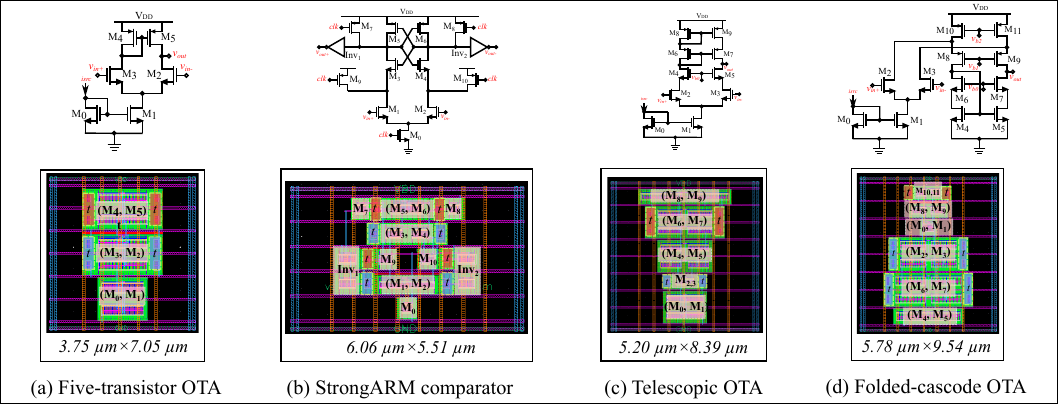}
    \caption{Schematics and corresponding layout realizations of the testcase circuits: (a)~5T OTA; (b)~StrongARM comparator; (c)~Telescopic OTA; (d)~Folded-cascode OTA. Red and blue overlays denote the optimized N-tap and P-tap regions.}
    \label{fig:result_figure}
\end{figure*}

\begin{figure}[t]
    \centering
\includegraphics[width=0.99\linewidth]{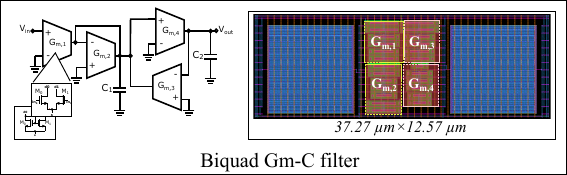}
    \caption{Biquad Gm-C filter schematic and corresponding quantized-height layout realization. Layout annotations indicate the individual Gm cells.}
    \label{fig:result_figure_gmc}
\end{figure}

Table~\ref{tab:benchmark_layout_summary} summarizes the physical layout results. Figure~\ref{fig:result_figure} and~\ref{fig:result_figure_gmc} show the synthesized QH layouts for the evaluated testcases with their corresponding schematics. Overall, the QH implementation remains area-competitive with the FH baseline while providing explicit row structure and regularized primitive organization. Although QH layouts can introduce additional whitespace
when devices are partitioned across multiple rows, the row-based QH fabric
also enables tighter primitive abutment and more systematic tap-aware
refinement. As a result, compared with the FH baseline, the proposed flow reduces layout area of the 5T~OTA, StrongARM comparator, folded-cascode OTA, and Gm-C filter by $21.2\%$, $24.1\%$, $2.41\%$, and $0.1\%$, respectively, whereas the telescopic OTA incurs an $8.8\%$ area overhead. In addition, the tap-aware refinement consistently reduces tap area across all testcases, with more than a $57\%$ reduction in the standalone OTA/comparator cases relative to the FH layout, and comparable tap area in the larger Gm-C filter. These results show that QH layout with tap optimization can improve regularity and tap efficiency while maintaining compact layout area in most cases, with the added advantage of reducing the S2S gap, as demonstrated in~\cite{Chou25}.
\input{tables/table_3_final_layout_summary}
\input{tables/table_4_ota_comparator_postlayout_comparison}
\input{tables/table_5_gmc_filter_fh_qh_comparison}

\subsubsection{Electrical Performance Comparison}
\label{subsubsec:electrical_comparison}
The postlayout electrical comparison is reported in Table~\ref{tab:elec_performance_ota_cmp} and Table~\ref{tab:gmc_filter_performance}. All postlayout results are obtained after Calibre parasitic extraction for both FH and QH implementations. Overall, the proposed QH implementation closely preserves the electrical behavior of the FH baseline well across the evaluated testcases. For the 5T OTA, the two layouts are nearly identical, with only $0.73\%$ degradation in unity-gain frequency (UGF), $0.42\%$ increase in power, and negligible changes in DC gain ($A_0$), phase margin (PM), and input-referred noise ($v_{n,\mathrm{in}}$). For the folded-cascode OTA, the QH version maintains UGF within $1.83\%$ of FH and improves PM by $7.6^\circ$, at the cost of moderate $A_0$ reduction and some degradation in the minimum slew-rate ($\text{SR}_{\min}$). The telescopic OTA also remains close to the FH baseline, with a $2.78\%$ UGF reduction, nearly unchanged power and noise, and improved $A_0$.
For the StrongARM comparator, the QH implementation incurs only an $8.8\%$ evaluation-delay~($t_{eval}$) overhead, a $5.07\%$ precharge-delay ($t_{pre}$) overhead, and a $3.40\%$  overhead in energy-per-comparison ($E_{\mathrm{cmp}}$). These results indicate that although the QH fabric imposes additional layout regularity constraints, it remains electrically competitive with the FH baselines for the standalone analog blocks, maintaining only single-digit percentage deviations across the key electrical metrics.

The Gm-C filter further evaluates this trend in a larger multi-primitive setting. Unlike the standalone OTA and comparator testcases, the filter requires multiple transconductance cells and capacitors to be arranged and routed across a wider layout region. In this setting, the proposed QH organization improves layout regularity, pin accessibility, and power-grid accessibility, reducing irregular routing detours compared with the FH implementation. This trend is consistent with the PDN accessibility study in Section~\ref{subsec:pg_benefit}, where the simplified DUT experiment shows the QH layout provides more direct and regular supply access. Because cutoff frequency ($f_c$) and passband gain ($A_{\mathrm{pass}}$) are target-matching metrics, the schematic-to-layout deviation is the most relevant comparison. As shown in Table~\ref{tab:gmc_filter_performance}, the FH layout shifts $f_c$ by $27.27\%$, whereas the QH layout limits this shift to $8.59\%$. This corresponds to a $68.5\%$ relative reduction in schematic-to-layout $f_c$ drift. Similarly, the FH $A_{\mathrm{pass}}$ changes by $-3.83$~dB after layout, while the QH implementation changes by only $+0.67$~dB. The QH layout also reduces power from $762.79~\mu$W to $664.86~\mu$W, corresponding to a $12.84\%$ reduction, with only a modest reduction in stopband attenuation ($A_{\mathrm{stop}}$). These results indicate that, for larger circuits composed of multiple analog primitives, the proposed QH fabric can reduce schematic-to-layout drift and improve implementation regularity while remaining electrically competitive with the FH baseline.

\subsection{Impact of QH Layout Regularity on Electrical Performance}
\subsubsection{Height Selection Study}
\label{subsec:height_selection}

\begin{figure}[t]
    \centering
\includegraphics[width=1\linewidth]{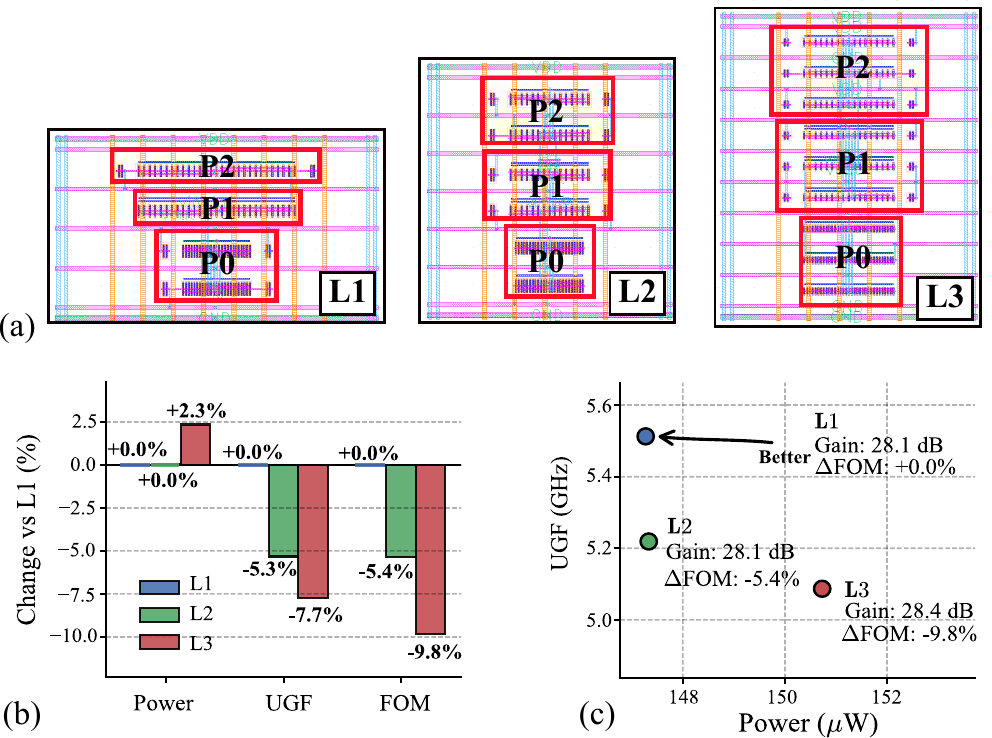}
    \caption{Height-selection and postlayout performance tradeoffs for three QH 5T OTA layouts: (a)~variants generated using different $\textit{fin}^{\mathrm{base}}$ values; (b)~comparison of power, UGF, and FOM with respect to layout L1; and (c)~corresponding power--UGF--$A_0$ tradeoff.}
\label{fig:optimized_height_comparison}
\end{figure}
Figure~\ref{fig:optimized_height_comparison} compares three quantized-height realizations, L1--L3, generated for a 5T OTA circuit with the same schematic target. The three layouts preserve nearly the same $A_0$, around $28.1$--$28.4$~dB; however, their bandwidth--power efficiency differs noticeably. Because the load capacitance is fixed across all cases, the OTA FOM~\cite{Sansen06} is proportional to $(\mathrm{UGF}/P)$. Relative to L1, L2 keeps nearly the same $P$ but reduces UGF by $5.3\%$, while L3 increases $P$ by $2.3\%$, reduces UGF by $7.7\%$, and lowers FOM by $9.8\%$. Therefore, the slightly higher $A_0$ of L3 does not offset its poorer bandwidth--power tradeoff. These tradeoffs can become more significant in larger multi-primitive circuits, where different height choices can lead to larger variations in interconnect length, pin accessibility, and device parasitic behavior. This suggests that the common quantized height has an electrical optimum, motivating the proposed MLP-based height-selection score, which selects L1 in this case.

\subsubsection{Local PDN Accessibility Benefits of QH Layout}
\label{subsec:pg_benefit}
\begin{figure}[t]
    \centering
    \includegraphics[width=1\linewidth]{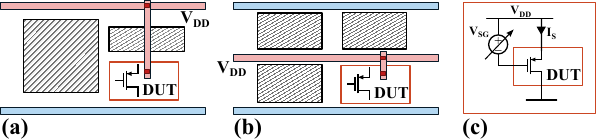}
\caption{Row-based layout method for improving PDN accessibility near the DUT: 
(a) Obstructed local $V_{DD}$ access in FH layout. 
(b) Nearby inner $V_{DD}$ access in the QH layout. 
(c) Equivalent DUT measurement circuit.}
    \label{fig:pdn_dut_access}
\end{figure}

\input{tables/table_6_pdn_accessibility_comparison}
To isolate the impact of layout style on local power delivery, we perform a controlled experiment using the same device under test (DUT) implemented in two layout scenarios: (i)~an FH layout, representing the arbitrary-height style used as the comparison baseline, and (ii)~a QH layout generated by the proposed approach. Figure~\ref{fig:pdn_dut_access} illustrates this practical situation, where local $V_{DD}$ access can become obstructed in large layouts with multiple horizontally placed primitives, as in the Gm-C filter testcase. In contrast, the QH layout provides nearby inner supply access to the DUT. 

For each layout scenario, the local supply-wire width is swept as an integer multiple of the minimum allowed wire width for the corresponding metal layer, denoted $1\times$, $2\times$, and $3\times$. As summarized in Table~\ref{tab:power_integrity_comparison}, the QH layout consistently improves the simulated DUT operating point compared with the FH layout. The QH implementation provides $g_m$ improvements of $7.0\%$, $6.3\%$, and $5.1\%$, with corresponding $I_D$ improvements of $5.2\%$, $4.7\%$, and $3.7\%$ for the $1\times$, $2\times$, and $3\times$ supply-width cases, respectively. These device-level improvements are important because local bias degradation can propagate to circuit-level metrics in large analog blocks. This mechanism is consistent with the Gm-C filter testcase discussed in Section~\ref{subsubsec:electrical_comparison}, where the QH implementation shows reduced schematic-to-layout drift compared with the FH baseline. The improvement trend is expected because the local lower-metal interconnects in advanced-node layouts are narrow and resistive, resulting in an IR voltage drop that degrades device operating conditions. By placing the local supply access closer to the DUT and reducing irregular supply-routing detours, the QH layout reduces this sensitivity. As the effective supply width increases from $1\times$ to $3\times$, the supply resistance is reduced, so the relative benefit of QH regularity becomes smaller, although it remains consistently positive. This supports the main motivation of the QH fabric: row-compatible layout regularity improves not only placement structure and routing access, but also the accessibility and robustness of local power delivery paths.

%% file: tables/table_3_final_layout_summary.tex
\begin{table}[t]
\centering
\caption{Final layout result summary.}
\label{tab:benchmark_layout_summary}
\scriptsize
\setlength{\tabcolsep}{2.0pt}
\renewcommand{\arraystretch}{0.95}
\begin{tabular}{|l|cc|r|cc|}
\hline
\textbf{Circuit} 
& \multicolumn{2}{c|}{\textbf{Layout/Taps:} \textbf{Area ($\mu \mathrm{m}^2$)}} 
& \textbf{$\Delta_{\mathrm{layout}}$} 
& \multicolumn{2}{c|}{\textbf{QH Structure}} \\ 
\cline{2-3}\cline{5-6}
& \textbf{FH}  
& \textbf{QH} 
& %\textbf{(\%)} 
& \textbf{Rows} 
& $\textit{fin}^{\textit{base}*}$ \\ 
\hline \hline

\makecell[l]{5T OTA}          
& $33.50/9.57$ & \textbf{$26.40/4.05$} & $-21.20\%$ 
& $6$  & $8$ \\ 
\hline

\makecell[l]{Telescopic OTA}  
& $40.10/10.61$ & $43.63/2.74$ & $+8.80\%$
& $8$  & $6$ \\ 
\hline

\makecell[l]{StrongARM comparator}           
& $44.00/10.20$ & \textbf{$33.40/3.65$} & $-24.09\%$ 
& $5$ & $6$ \\ 
\hline

\makecell[l]{Folded-cascode OTA}
& $56.50/12.69$ & \textbf{$55.14/5.32$} & $-2.41\%$  
& $8$  & $9$ \\ 
\hline

\makecell[l]{Gm-C filter}     
& $469.30/21.50$ & \textbf{$468.50/20.68$} & $-0.10\%$      
& $13$ & $5$ \\ 
\hline
\end{tabular}
\end{table}

%% file: tables/table_4_ota_comparator_postlayout_comparison.tex
\begin{table*}[t]
% \vspace{-0.4cm}
\centering
\caption{Post-layout comparison of free-height (FH) and quantized-height (QH) implementations for OTA and comparator testcases.}
\label{tab:elec_performance_ota_cmp}
\resizebox{\linewidth}{!}{
\setlength{\tabcolsep}{5pt}
\renewcommand{\arraystretch}{1}
\begin{tabular}{|l|r|r|r|r|r|r|r|r|r||l|r|r|r|}
\hline
\multicolumn{10}{|c||}{\textbf{OTAs}} 
& \multicolumn{4}{c|}{\textbf{Comparator}} \\
\hline
\textbf{Metric} & \multicolumn{3}{c|}{
\begin{tabular}[c]{@{}c@{}}
\textbf{5T OTA}\\
\hline
\footnotesize ${N_{\text{dev}}=6, N_{\text{prim}}=3}$
\end{tabular}
}
& \multicolumn{3}{c|}{
\begin{tabular}[c]{@{}c@{}}
\textbf{Telescopic OTA}\\
\hline
\footnotesize ${N_{\text{dev}}=10, N_{\text{prim}}=5}$
\end{tabular}
}

& \multicolumn{3}{c||}{
\begin{tabular}[c]{@{}c@{}}
\textbf{Folded-cascode OTA}\\
\hline
\footnotesize ${N_{\text{dev}}=12, N_{\text{prim}}=6}$
\end{tabular}
}

& \textbf{Metric} & 
\multicolumn{3}{c|}{
\begin{tabular}[c]{@{}c@{}}
\textbf{StrongARM comparator}\\
\hline
\footnotesize ${N_{\text{dev}}=15, N_{\text{prim}}=10}$
\end{tabular}
}\\

\cline{2-10}\cline{12-14}
& \textbf{FH} & \textbf{QH} & $\Delta$
& \textbf{FH} & \textbf{QH} & $\Delta$
& \textbf{FH} & \textbf{QH} & $\Delta$
& & \textbf{FH} & \textbf{QH} & $\Delta$ \\
\hline\hline

$A_0$ (dB)
& $30.64$ & $30.65$ & $+0.01$~dB
& $37.46$ & $39.19$ &  $+1.73$~dB
& $44.55$ & $41.05$ & $-3.5$~dB
& \multirow{2}{*}{\begin{tabular}[l]{@{}l@{}} $t_{eval}$ (ps)\end{tabular}}
& \multirow{2}{*}{$144.3$} 
& \multirow{2}{*}{$157.0$} 
& \multirow{2}{*}{$+8.80\%$} \\
\cline{1-10}

UGF (MHz)
& $204.60$ & $203.10$ & $-0.73\%$
& $763.40$ & $734.53$ & $-2.78\%$
& $207.70$ & $203.90$ & $-1.83\%$
& & & & \\
\hline

PM ($^\circ$)
& $79.31$ & $79.40$ & $+0.09^\circ$
& $67.01$ & $65.70$ & $-1.31^\circ$
& $74.80$ & $82.40$ & $+7.60^\circ$
& \multirow{2}{*}{\begin{tabular}[l]{@{}l@{}}$t_{pre}$ (ps)\end{tabular}}
& \multirow{2}{*}{$73.00$} 
& \multirow{2}{*}{$76.70$} 
& \multirow{2}{*}{$+5.07\%$} \\
\cline{1-10}

SR (V/$\mu$s)
& $715.20$ & $695.80$ & $-2.71\%$
& $434.90$ & $510.04$ & $+17.28\%$
& $129.47$ & $117.01$ & $-9.63\%$ 
& & & & \\
\hline

Power ($\mu$W)
& $9.62$ & $9.66$ & $+0.42\%$
& $15.80$ & $15.92$ & $+0.76\%$
& $32.90$ & $33.44$ & $+1.64\%$
& E$_{\mathrm{cmp}}$ (fJ) & $150.66$ & $155.79$ & $+3.40\%$ \\
\hline

$v_{n,\mathrm{in}}$ (nV/$\sqrt{\mathrm{Hz}}$)
& $24.73$ & $24.75$ & $+0.08\%$
& $43.59$ & $43.57$ & $-0.05\%$
& $47.38$ & $50.96$ & $+7.56\%$
& \multicolumn{4}{c}{} \\
\cline{1-10}
\multicolumn{14}{l}{\footnotesize $N_\mathrm{dev}$: number of devices; $N_\mathrm{prim}$: number of generated analog primitives.}
\end{tabular}
}
\end{table*}

%% file: tables/table_5_gmc_filter_fh_qh_comparison.tex
\begin{table}[t]
\centering
\caption{FH vs. QH implementations for the Biquad Gm-C filter.}
\label{tab:gmc_filter_performance}
\scriptsize
\setlength{\tabcolsep}{2.5pt}
\renewcommand{\arraystretch}{1.2}
\begin{tabular}{|l|c|c|c|c|c|c|}
\hline
\multicolumn{7}{|c|}{
\begin{tabular}[c]{@{}c@{}}
\textbf{Biquad Gm-C Filter}\\
\hline
\footnotesize ${N_{\text{dev}}=26, N_{\text{prim}}=14}$
\end{tabular}
}\\
\hline
\multirow{2}{*}{\textbf{Metric}} 
& \multicolumn{3}{c|}{\textbf{FH}} 
& \multicolumn{3}{c|}{\textbf{QH}} \\
\cline{2-7}
& \textbf{Schematic} & \textbf{Layout} & $\Delta$
& \textbf{Schematic} & \textbf{Layout} & $\Delta$ \\
\hline\hline
$f_c$ (MHz) 
& 293.94 & 374.10 & +27.27\%
& 292.84 & 318.00 & +8.59\% \\
$A_{\mathrm{pass}}$ (dB) 
& 1.82 & -2.01 & -3.83 dB
& -0.85 & -0.18 & +0.67 dB \\
\hline
& \multicolumn{2}{c|}{\textbf{FH}} 
& \multicolumn{2}{c|}{\textbf{QH}} 
& \multicolumn{2}{c|}{$\Delta$} \\
\hline
$A_{\mathrm{stop}}$ (dB) 
& \multicolumn{2}{c|}{$20.10$} 
& \multicolumn{2}{c|}{$18.94$} 
& \multicolumn{2}{c|}{$-1.16$ dB} \\
Power ($\mu$W) 
& \multicolumn{2}{c|}{$762.79$} 
& \multicolumn{2}{c|}{$664.86$} 
& \multicolumn{2}{c|}{$-12.84 \%$} \\
\hline
\end{tabular}
% \vspace{-0.5cm}
\end{table}

% \begin{table}[h!]
% \centering
% \caption{Comparison of FH and QH implementations for the Biquad Gm-C filter.}
% \label{tab:gmc_filter_performance}
% \scriptsize
% \setlength{\tabcolsep}{2.5pt}
% \renewcommand{\arraystretch}{1.2}
% \begin{tabular}{|l|c|c|c|c|c|c|}
% \hline
% \multicolumn{7}{|c|}{
% \begin{tabular}[c]{@{}c@{}}
% \textbf{Biquad Gm-C Filter}\\
% \hline
% \footnotesize ${N_{\text{dev}}=26, N_{\text{prim}}=14}$
% \end{tabular}
% }\\
% \hline
% \multirow{2}{*}{\textbf{Metric}} 
% & \multicolumn{3}{c|}{\textbf{FH}} 
% & \multicolumn{3}{c|}{\textbf{QH}} \\
% \cline{2-7}
% & \textbf{Sch.} & \textbf{Layout} & $\Delta$
% & \textbf{Sch.} & \textbf{Layout} & $\Delta$ \\
% \hline\hline
% $f_c$ (MHz) 
% & 293.94 & 374.10 & +27.27\%
% & 292.84 & 318.00 & +8.59\% \\
% \hline
% & \multicolumn{2}{c|}{\textbf{FH}} 
% & \multicolumn{2}{c|}{\textbf{QH}} 
% & \multicolumn{2}{c|}{$\Delta(\%)$} \\
% \hline
% Passband gain (dB) 
% & -2.01 & -0.18 &   \\????????? (still doesn't make sense to compare the two results for this metrics)
% Stopband attenuation (dB) 
% & \multicolumn{2}{c|}{20.10} 
% & \multicolumn{2}{c|}{18.94} 
% & \multicolumn{2}{c|}{-5.77} \\
% Power ($\mu$W) 
% & \multicolumn{2}{c|}{762.79} 
% & \multicolumn{2}{c|}{664.86} 
% & \multicolumn{2}{c|}{-12.84\%} \\
% \hline
% \end{tabular}
% \vspace{-0.5cm}
% \end{table}

%% file: tables/table_6_pdn_accessibility_comparison.tex
\begin{table}[t]
\centering
\caption{Performance Comparison: Free vs. Row-Based Layouts.}
\label{tab:power_integrity_comparison}
\scriptsize
\setlength{\tabcolsep}{4pt} 
\begin{tabular}{|l|l|c|c|c|}
\hline
\textbf{Wire} & \textbf{Layout} & \boldmath{$g_m$}  & \boldmath{$I_D$}  & \textbf{QH vs. FH} \\ 
&   & \textbf{(mS)} & \textbf{($\mu$A)} & \textbf{Improvement} \\ \hline \hline
$1\times$ & FH & $4.87$ & $731$ & $g_m$: $+7.0\%$ \\ \cline{2-4} 
& QH & $5.21$ & $769$ & $I_d$: $+5.2\%$ \\ \hline \hline
$2\times$ & FH & $5.72$ & $824$ & $g_m$: $+6.3\%$ \\ \cline{2-4}
& QH & $6.08$ & $862$ & $I_d$: $+4.7\%$ \\ \hline \hline
$3\times$ & FH & $6.11$ & $865$ & $g_m$: +5.1\% \\ \cline{2-4}
& QH & $6.42$ & $898$ & $I_d$: $+3.7$\% \\ \hline
\end{tabular}
\end{table}

%% file: contents/6_Conclusion.tex
\section{Conclusion}
\label{sec:conclusion}
\noindent
This work introduced a height-quantized, row-based layout synthesis methodology for FinFET analog and mixed-signal circuits, which has previously been shown to reduce the S2S gap~\cite{Chou25}. By integrating  fixed-height primitive realization, and custom well- and tap-aware placement, the flow generates structured analog layouts with standard-cell-like regularity while preserving key analog design constraints. The results demonstrate that the approach retains the advantages of regular layout, with performance comparable to free-height designs. 